\documentclass[12pt,preprint]{aastex}

\slugcomment{Accepted for Publication in the Astrophysical Journal}

\shorttitle{Variability in HD 31648 and HD 163296}
\shortauthors{Sitko et al.}

\begin{document}


\title{Variability of Disk Emission in Pre-Main Sequence  and Related Stars. \\
    I. HD 31648 and HD 163296 - Isolated Herbig Ae Stars Driving Herbig-Haro Flows}


\author{Michael L. Sitko\altaffilmark{1,2}, William J. Carpenter, Robin L. Kimes, J. Leon Wilde\altaffilmark{3}}
\affil{Department of Physics, University of Cincinnati, Cincinnati OH 45221}
\email{sitko@physics.uc.edu, carpenwj@physics.uc.edu, kimesrl@email.uc.edu, jlwilde@bgnet.bgsu.edu}

\author{David K. Lynch\altaffilmark{1}, Ray W. Russell\altaffilmark{1}, Richard J. Rudy, Stephan M. Mazuk, and Catherine C. Venturini}
\affil{The Aerospace Corporation, Los Angeles, CA 90009}
\email{David.K.Lynch@aero.org, Ray.W.Russell@aero.org, Richard.J.Rudy@aero.org, Stephan.M.Mazuk@aero.org, Catherine.C.Venturini@aero.org}

\author{Richard C. Puetter}
\affil{Center for Astrophysics and Space Science, University of California, San Diego, CA 92093}
\email{rpuetter@ucsd.edu}

\author{Carol A. Grady}
\affil{Eureka Scientific, Inc., Oakland, CA 94602 \\ and \\ Exoplanets and Stellar Astrophysics Laboratory, \\ Code 667, Goddard Space Flight Center, Greenbelt, MD 20771}
\email{cgrady@milkyway.gsfc.nasa.gov}

\author{Elisha F. Polomski\altaffilmark{4}}
\affil{Department of Astronomy, University of Minnesota, Minneapolis, MN 55455}
\email{Elisha.Polomski@uwsp.edu}

\author{John P. Wisnewski}
\affil{Exoplanets and Stellar Astrophysics Laboratory, \\ Code 667, Goddard Space Flight Center, Greenbelt, MD 20771}
\email{jwisnie@miklyway.gsfc.nasa.gov}

\author{Suellen M. Brafford}
\affil{1368 Gumbert Dr., Amelia, OH 45102}
\email{SMBlawyer@fuse.net}

\author{H. B. Hammel}
\affil{Space Science Institute, 4750 Walnut Street, Suite 205, Boulder, CO 80301}
\email{hbh@alum.mit.edu}

\and

\author{R. Brad Perry}
\affil{NASA Langley Research Center, Hampton, VA 23681}
\email{Raleigh.B.Perry@NASA.gov}

\altaffiltext{1}{Visiting Astronomer, NASA Infrared Telescope Facility, operated by the University of Hawaii under contract with the National Aeronautics and Space Administration.}
\altaffiltext{2}{also Space Science Institute}
\altaffiltext{3}{now at Solar One}
\altaffiltext{3}{now at the University of Wisconsin, Stevens Point}




\begin{abstract}
Infrared photometry and spectroscopy covering a time span of a quarter century are presented for HD 31648 (MWC 480) and HD 163296 (MWC 275). Both are isolated Herbig Ae stars that exhibit signs of active accretion, including driving bipolar flows with embedded Herbig-Haro (HH) objects. HD 163296 was found to be relatively quiescent photometrically in its inner disk region, with the exception of a major increase in emitted flux in a broad wavelength region centered near 3 $\mu$m in 2002. In contrast, HD 31648 has exhibited sporadic changes in the entire 3-13 $\mu$m region throughout this span of time. In both stars the changes in the 1-5 $\mu$m flux indicate structural changes in the region of the disk near the dust sublimation zone, possibly causing its distance from the star to vary with time. Repeated thermal cycling through this region will result in the preferential survival of large grains, and an increase in the degree of crystallinity. The variability observed in these objects has important consequences for the interpretation of other types of observations. For example, source variability will compromise models based on interferometry measurements unless the interferometry observations are accompanied by nearly-simultaneous photometric data.

\end{abstract}

\keywords{circumstellar matter - infrared:stars - planetary systems: protoplanetary disks - stars:formation - stars: pre-main-sequence - stars:individual (HD 31648, HD 163296)}



\section{Introduction}

HD 31648 (MWC 480) and HD 163296 (MWC 275) are  ``isolated'' Herbig Ae/Be stars (HAeBes) that bridge the gap between classical pre-main sequence (PMS) and main sequence (MS) stars  in the sense that, although they are Ae stars that lie above the MS and exhibit emission lines and infrared excesses, they are not immediately associated with dark clouds or reflection nebulae.

With a distance of 122$^{+17}_{-13}$ pc and an age of 4$^{+6}_{-2.5}$ Myr \citep{vda98}, HD 163296 appears to be an outlying member of the Upper Cen-Sco OB association. Its spectral energy distribution (SED) at wavelengths shortward of 15 $\mu$m is very similar to that of the classical HAeBe star AB Aur \citep{sit81} with a strong continuum infrared excess and 10-$\mu$m silicate emission band. HD 163296 is often considered a prototype of the ``isolated'' HAeBe star class. At longer wavelengths, the SEDs differ: HD 163296 lacks the extended ``plateau'' exhibited by AB Aur between $\lambda \sim$ 20 - 60 $\mu$m \citep{me01}. Recent coronagraphic observations show the presence of an underlying disk of dust, and outflowing gaseous jets in Ly$\alpha$ \citep{dev00} that include a string of Herbig-Haro (HH) objects \citep{gra00}. At visible wavelengths, the dusty disk can be traced to 450 AU (3.7 arcsec)  \citep{gra00}. Its inclination has has been reported to be between 45$^ {\circ}$ and 65$^ {\circ}$ \citep{ms97,gra00,was06,ise07}, and we have adopted a value of 50$^{\circ}$ for our modeling. It also exhibits both far-UV emission lines \citep{del05} and X-ray emission that have been attributed to optically-thin shock-heated gas  that is accreting onto the stellar surface, most likely along magnetic field lines \citep{swa05}. 

HD 31648 has a distance based on \textit{Hipparcos} measurements of 131$^{+24}_{-18}$ pc and age of 2.5$^{+1.5}_{-1.0}$ Myr \citep{vda98}, which would place it at the edge of the Tau-Aur T association. Using this same distance, \citet{blo06} derive an age of 9 Myr. However, \citet{sim01} suggest a distance of 170 pc and an age of $\sim$7 Myr. It also exhibits an outflow of HH objects \citep{ste07} and emission lines of Si IV \citep{ssm81} consistent with hot accreting gas.  Its dusty disk has recently been detected in scattered visible light \citep{gra08}, but it is faint and has a dropoff in surface brightness with radial distance that is steeper than most systems detected so far. The dust continuum is also detected 1.3 mm \citep{ms97}. 

For both objects, the probable ages correspond roughly to the time when grain growth into larger bodies (i.e. cometesimals) has been operating, and in some models \citep{boss97} planetary embryos and perhaps true planets have already formed.

Analysis of the infrared SEDs of Herbig Ae stars obtained with the {\it Infrared Space Observatory (ISO)} show that most of these fall into one of two broad groups - those whose infrared emission that can be fitted (approximately) with simple power-laws, and those that require the addition of another roughly  ``blackbody'' component  \citep{me01,avda04}. These form the Group II and Group I sources in the nomenclature of \citet{me01}. \citet{dd04} suggest that these represent an evolutionary scenario, where Group II objects began originally as Group I objects with flared disks irradiated by the star, but as the disks age, grain growth decreases the optical depth of the disk, which then flattens and is eventually shadowed by a ``puffed-up'' inner rim of the disk.  This inflated inner disk wall results when a paucity of accreting gas inside the dust sublimation zone (hereafter DSZ)  leads to direct illumination of the dust  by the star, causing the location of the transition zone to migrate outward and to expand vertically, compared to one that is shadowed by the accreting gas [see Fig. 2 of \citet{mg07} for example]. 

Eventually, inner disk clearing, which can occur beyond the dust sublimation radius, will give rise to a third class of SEDs: those where the inner region is cleared due to the cessation of accretion, as well as the possible clearing by objects of planetary mass. \citet{cal02} and \citet{ber04} have recently provided strong evidence for the clearing of the inner disks of a number of PMS stars. In the disks surrounding TW Hya, GM Aur, DM Tau, and LkCa15, the spectral energy distributions require substantial clearing of material for distances closer to the star than a few AU. This process gives rise to the recently-recognized ``transition disk'' objects. \citet{gra07} find similar evidence for the star HD 169142, another such ÒtransitionÓ object with a substantial dust wall near 30 AU, and an inner disk with a tremendous paucity of dust.  In fact, many objects in the original Meeus Group I are not young objects with flared disks, but actually stars with transition disks with thick cool disks with large inner radii.

The SEDs of HD 31648 and HD 163296 are consistent with them being members of the Meeus Group II. HD 163296 had the faintest disk of all of the Ae stars in the sample of \citet{gra05}, with a middle zone that is darker than  the outermost traceable reaches of the disk, possibly signaling its partial shadowing by the inner disk. The faintness and radial dependence of the surface brightness  in the disk of HD 31648 \citep{gra08} are also consistent with significant dust  settling and shadowing.

It is surprising, then, that both HD 31648 and HD 163296  exhibit many indicators of the active accretion generally associated with flared unshadowed Group I objects: hot non-stellar gas, jets, and HH objects. In fact, of the 4 HAeBes with known HH knots (HD 163296, HD 31648, HD 104237, and AB Aur), the first three are Group II sources, in seeming contradiction to the general concept hat these Group II objects result from very low accretion rates. Apparently, the accreting gas can in some instances does not shadow the inner dust wall sufficiently, so that gas accretion and shadowing are not mutually exclusive. This might occur if the inner gas disk has a small scale height compared to the DSZ, or if it becomes optically thin in the radial direction. The degree to which the inner gaseous accretion disk can actually shadow the dust further out is not known precisely, but \citet{muz04} suggest that the gaseous disks becomes optically thick radially once the accretion rate drops below a few times 10$^{-8}$ M$_{\sun}$ yr$^{-1}$. Its vertical scale height is sufficient to shadow the dust disk midplane, but not the entire inner wall of the dust disk.

Highly collimated outflows require an efficient collimating mechanism. In the T Tauri stars this mechanism is generally thought to result from magnetic fields in the star, possibly connected to the disk \citep{shu94}.  \citet{hu06} reported the detection of circular polarization in the circumstellar Ca II lines, but the implied derived field strength in this outflowing material is uncertain, as the polarization detection was only at a 2-$\sigma$ level. No photospheric magnetic field was detected. They  reported a 4-$\sigma$ detection of a magnetic field in HD 31648, also mostly of circumstellar origin. However, \citet{wade07}, with comparable observational uncertainties,  found no detectable field. Only further study will reveal if  these  stars have significant, but variable, fields.

The spacing and proper motions of the HH knots in HD 163296 \citep{gra00,swa05} requires changes in ejection activity on time scales of about 5-6 years. If these are controlled by the inner disks (inner gas disk or the gas+dust disk further out), then changes in disk structure may fuel the outbursts. On the other hand, if the outbursts are driven by the physics of a magnetic field (such as reconnection events) rather than the fuel supply, then outburst activity might affect the inner disk structure through increased far-UV and X-ray activity and changes in disk winds generated by the activity. Either way, one might expect changes in the innermost regions of the disk in such objects on time scales that are similar to that of the generation of new HH objects.

As part of a long-term program to investigate the mid-IR ($\sim$3-13$\mu$m) brightness variations of HAeBe and related stars, we have observed HD 31648 and HD 163296 at these wavelengths on multiple epochs. Here we describe the nature of the observed variations, and how these relate to disk structure and accretion activity. The complete sample of objects observed as part of this program will be discussed in a later paper.

\section{Observations}

Observations were carried out over a span of a quarter century using a variety of photometric and spectroscopic instruments. Table 1 lists the dates of the observations and the instruments used in each case.

The bulk of the data presented here that cover the 3-13 $\mu$m region were obtained between 1994 and 2006 using the Aerospace Corporation's Broad-band Array Spectrograph System (BASS) on NASA's Infrared Telescope Facility (IRTF). The 2004 BASS data were obtained at the Mount Lemmon Observing Facility (MLOF). This instrument uses a cold beamsplitter to separate the light into two separate wavelength regimes. The short-wavelength beam includes light from 2.9-6 $\mu$m, while the long-wavelength beam covers 6-13.5 $\mu$m. Each beam is dispersed onto a 58-element Blocked Impurity Band (BIB) linear array, thus allowing for simultaneous coverage of the spectrum from 2.9-13.5 $\mu$m. The spectral resolution $R = \lambda$/$\Delta\lambda$ is wavelength-dependent, ranging from about 30 to 125 over each of the two wavelength regions. The simultaneous coverage of the entire 3-13 $\mu$m region that BASS provides  has played a critical role in the investigations of the clearing of the inner disks in systems like those we are discussing here \citep{cal02,ber04}.

At the IRTF, the circular entrance aperture of the instrument subtends 3.4 arcsec on the sky (8.5 arcsec at the MLOF) . At the distance of HD 163296 the BASS/IRTF 1.7-arcsec radius corresponds to approximately 210 AU, while the disk can be traced to 450 AU in coronagraphic images \citep{gra00}. However, the disk has  a half-width at half-maximum brightness of 0.007 arcsec at 12.5 $\mu$m \citep{lei04}, and a half-width half-maximum  of less than 0.5 arcsec  at 1.3 mm wavelength \citep{ms97}. In this paper we will limit our discussions to wavelengths of 20 $\mu$m and shorter, since the entrance aperture includes essentially all the radiation emitted at wavelengths shortward of 20 $\mu$m. A wavelength of 20 $\mu$m corresponds (in units of $\lambda F_{\lambda}$    $W m^{2}$) to a characteristic blackbody temperature of $T\sim180 K$, the equilibrium temperature at 12 AU (0.1 arcsec for a distance of 122 pc) from a star with $L\sim 26 L_{\sun} $ \citep{vda98}. 

Both the 1.3 mm interferometry   \citep{ms97} and coronagraphic scattered light imaging \citep{gra00,ste07} of HD 31648 indicate that it has an even smaller disk than HD 163296.  Because all of the other entrance apertures used in this study were larger than this value, the flux at the wavelengths sampled for variability (i.e., less than 20 $\mu$m), will be contained within all of the entrance apertures used.

Prior to 1996, the data on both stars were obtained using primarily medium-bandpass filters with spectral resolutions of $R \sim 6$,  and  narrower-bandpass filters with $R \sim 50$ near the 10-$\mu$m silicate band. When searching for possible variability in the infrared emission, it is critically important to insure that all the observations are based on the same relative and absolute flux systems. Unfortunately, throughout the history of infrared astronomy, a wide variety of filters, standard star magnitude systems, and absolute flux calibrations have been in use.  For this study, we have applied a single relative brightness and absolute flux calibration system, described more fully in  Appendix A.

High (R$\sim$1000) spectral resolution measurements at shorter wavelengths  were carried out in 2002 and 2005 using The Aerospace Corporation'Õs Visible and Near-Infrared Imaging Spectrograph (VNIRIS).  The original infrared portion of the spectrograph (NIRIS) is described by \citet{ru99}. It employs two independent channels separated by a beamsplitter covering the wavelength ranges 0.8-1.4 $\mu$m and 1.4-2.5 $\mu$m, respectively.  Each channel employs its own collimator, camera, grating, and Hawaii 1 HgCdTe detector array, but views the slit through a common field lens.  This was the configuration for the 2002 observations of HD 163296.  By 2005, a second beamsplitter and a  third channel, covering the 0.4-0.9 $\mu$m, had been added to the system, now called VNIRIS (a term which we will use for both versions throughout the remainder of this paper), and used to observe the same star that year.  It resides outside of the cryogenic vessel but also views the same slit through the same field lens as the infrared channels.  The detector in this case is a red-sensitive, deep-depletion CCD.  

Dispersion is nearly uniform over the wavelengths covered by each channel of the VNIRIS spectrograph, implying that resolution the $R = \lambda$/$\Delta\lambda$ increases linearly with wavelength.  For objects that underfill the slit width, resolution is determined by seeing and guiding.  We measured the resolution directly using the unresolved emission lines of HD 163296.  For the July 2002 observations it ranged from approximately 550 to 950 for the 0.8-1.4 $\mu$m wavelength range spanned by the short wavelength channel; for the long wavelength channel, the resolution varied from about 500 at 1.4 $\mu$m to 900 at 2.5 $\mu$m.  For the 2005 observations the resolution for the infrared channels was very similar to that in 2002.  For the optical, which did not exist for the July 2002 observations, the resolution ranged from 600 to 1000 between 0.5 $\mu$m and 0.9 $\mu$m.  Data shortward of 0.5 $\mu$m were acquired in second order; the resolution varied from 900 to 1150.

The July 2005 observations used the solar type star HD 162255 for a standard.  For a model of its spectrum, we used that of the sun and modified the shapes and strengths of the stronger absorption lines (e.g., H$\alpha$, the infrared Ca II triplet) to match the actual observations of the standard.  The same process was applied to the earlier type standard used for the July 2002 observations, but for it the main focus was the H I lines.

\section{Results}

The SEDs of HD 163296 and HD 31648, from ultraviolet to millimeter wavelengths, are illustrated in Figures 1 and 2. For both objects, the flux in units of $\lambda$F$_{\lambda}$ $W m^{-2}$ is lower at 60 $\mu$m than at 8 $\mu$m, consistent with a Group II classification. Neither object exhibits significant variation in the photometric V band at 0.55 $\mu$m: historically these have been restricted to being less than 20\% for HD 31648 and 10\% for HD 163296 \citep{dolf01}. In the near-infrared (JHK bands), the variations are larger for HD 163296 than in the V band, being over 30\% at J(1.2 $\mu$m), 20\% at H(1.6 $\mu$m), and 15\% at K(2.2 $\mu$m) \citep{dolf01}.  HD 31648 has not been extensively monitored, and little information is available from the literature on its near-IR variability, but based on our limited data set, HD 31648 seems to exhibit near-IR variability at least as large as that of HD 163296.

\subsection{HD 163296}

For HD 163296, Figure 3 shows data obtained on 14 dates, where we have supplemented our observations with select data from the literature. Many sets of these observations were obtained within a few weeks of one another, and together could be considered single  ``epochs'', providing coverage on a total of 10 epochs.  Observations by our team began in 1979 with the Kitt Peak bolometer, and after a hiatus of 17 years, resumed with BASS in 1996.  Beginning in 1996,  we obtained  BASS spectra on 6 separate epochs. On two of these, 2002 and 2005, we also observed the star with VNIRIS within three weeks of the BASS observations. In addition, the Carlos S\'{a}nchez telescope of the Teide Observatory on Tenerife obtained JHK data \citep{eir01} on this star within 4 days of the BASS observations in 1998. The \textit{Infrared Space Observatory} (\textit{ISO})  Short Wavelength Spectrograph (SWS) data in Figure 3 were obtained within 4 days of the 1996 BASS spectrum as well, and are in very good agreement at all wavelengths in common. We have also included three epochs of JHKLM photometry from \citet{dolf01}  between 1980 and 1986. 

Most of the data are consistent with no variability in excess of 10\%. In 2002, however, there was a very significant rise in the flux level between 1 and 5 $\mu$m, with perhaps only a slight increase in flux in the 10 $\mu$m silicate feature. We were fortunate to have VNIRIS and BASS observations in both low-flux  (2005) and high-flux states (2002). The combined VNIRIS-BASS data for the two epochs are shown in Figure 4. No shifts in flux level has been applied to these observations in order to make them ``match''  at any wavelength. The joint VNIRIS-BASS spectra clearly show that the ``outburst'' in 2002 was seen in both instruments at both telescopes, and is unlikely to be the result of instrumental effects and data processing.

The fairly regular spacing of the HH objects associated with the outflow jet in HD 163296 requires  quasi-periodic  ejection events roughly every 5-6 years.  \citet{dolf01} report JHKLM photometry during three epochs in the gap on our own coverage between 1979 and 1996 (labeled as ``ESO'' in Figure 3). Figure 5 illustrates the changing brightness of the 1-5 $\mu$m emission as measured at 3.77 $\mu$m using the ESO and BASS data (the 1979 KPNO observations were at 3.5 $\mu$m and of lower quality, and have been omitted here). Two of the ESO points (1980 and 1983)  are clearly consistent with the low flux state seen in most epochs, while their 1986 observations seem to have been obtained in another earlier outburst state. The data seems consistent with a variability time scale of about 16 years, although shorter time scales cannot be ruled out due to the paucity of the time coverage. There is also insufficient temporal coverage to decide if the changes are periodic. Hence, we cannot yet tell for sure if the variations observed are related to the quasi-periodic ejection of the HH objects.

\subsection{HD 31648}

For HD 31648, 10 epochs of data are shown in Figure 6. Our temporal coverage for HD 31648 is less extensive than for HD 163296. We include JHK photometry from the 2MASS survey and data obtained with the  \textit{ISO} SWS. Of the three sets of BASS data that simultaneously cover the 3-13 $\mu$m region, none shows evidence of a truly strong, well-defined 1-5 $\mu$m hump (except perhaps in its highest flux state) that is the hallmark of objects whose SEDs in that spectral region are interpreted as being due to the irradiated puffed-up inner disk wall, and which is usually used to invoke the self-shadowing picture of the Meeus Group II sources. As will be discussed below, the weakness of this feature may also be inconsistent with the interpretation of visible-wavelength interferometry measurements, which generally predict a substantially stronger feature \citep{mon06,ise06}. 

Unlike HD 163296, there is yet no indication of a ``preferred'' lower flux state.  The observed variability in HD 31648 is also of greater relative amplitude than that observed so far in HD 163296, and is more prominent in the 10 $\mu$m silicate band, as shown in Figure 4.  Figure 7 shows the BASS-derived fluxes at 3.77 $\mu$m and 10.5 $\mu$m for both stars. For HD 31648, the flux in the 3 $\mu$m and 10 $\mu$m regions vary together, whereas the only significant change in the 10 $\mu$m emission of HD 163296 occurred in the 2002 outburst. Nevertheless, the similarity in the ``difference'' spectra indicate that we are seeing the same basic phenomenon on both objects.

\subsection{Monte Carlo Radiative Transfer Models of the Disks}

In order to better understand the nature of the dusty disks surrounding HD 31648 and HD 163296, we have used of the Monte Carlo radiative transfer code of \citet{whi03a,whi03b,whi04} to model the SEDs of these systems. The code is a simulation using many photons propagating through, and interacting with, a circumstellar environment to produce the SED. The circumstellar dust model is separated into three physical zones: a disk, an envelope, and an outflow region.  Each region is assigned user-defined geometrical dimensions and grain properties. For the disk, the inner and outer radii, inner edge disk scale height, mass, and the amount of flaring can be set. When hydrostatic equilibrium and mass conservation in the accretion flow are enforced, the radial mass density and scale height are coupled. In the case of an irradiated $\alpha$-type disk, the surface density $\Sigma \propto R^{-1}$ \citep{paola98} where R is the radial distance. For a radial mass density $\rho \propto R^{-A}$ and density scale height $H \propto R^{B}$, mass conservation in the accretion flow requires $A - B = 1$  \citep{rob06}, a restriction that was imposed in these models (see Table 2).   Changing $B$ alters the degree of flaring, while maintaining the restrictions of hydrostatic equilibrium and conservation of mass accretion at any given radial distance. $B = 1$ describes an unflared disk (i.e. conical opening), while $B  > 1$ is for a standard ``concave upward'' flared disk. $B < 1$ can be used for an ``anti-flared'' convex upward disk. The last of these may be appropriate for disks with grain settling. Envelope properties include inner and outer radii and the rate of flow of material onto the star. Such an envelope can also be used to simulate a disk wind or mini-halo.  The outflow region is useful for including the effects of jets, as is observed in many young disk systems. This dust model is then overlaid on a radial/azimuthal grid, and the material density of each region is calculated for every grid cell. The main disk itself is a blend of two regions, the disk midplane and an upper disk region analogous to a disk ``atmosphere'' discussed by many investigators. Transition between grain characteristics going from midplane to the disk surface can be implemented when the mass density crosses a specified threshold, which in our model of HD 163296 was H$_{2}$ density $= 10^{10} cm^{-3}$, and larger sized dust grains were assigned to the disk midplane than the atmosphere to simulate the effects of dust settling. For HD 31648, the grain size was uniform throughout the disk.

All of the dust property files used here consist of nearly half and half mixture of amorphous carbon and silicates. Three dust files (of the four that come with the standard release of the radiative transfer code) with differing grain size distributions are used in the modeling, although it is not necessary (or desirable) to use them all in every object studied.  The first is the interstellar medium (ISM) grain model of \citet{kmh94} (hereafter KMH), which utilizes a power law size distribution with an exponential cutoff. The second one, built on the KMH model, is the one used by \citet{cot01} for modeling the scattered light in the upper disk atmosphere of HH 30. This one has a flatter size distribution, and a mean grain size about twice that of KMH - the size where the distribution has a turnover is near 0.55 $\mu$m, as opposed to $\sim$ 0.2 $\mu$m for KMH. Finally, a ``large'' grain model of \citet{wood02} for the infrared SED of HH 30 was used, which is similar to the other two, but with the turnover near 50 $\mu$m. In each region, the grain temperature is determined by the locally-generated accretion, irradiation from the star, and re-radiation from within the disk. For simplicity, a single temperature is used in each cell, based on the mean grain properties \citep{bw01}. In the fitting process, the SED in the millimeter-submillimeter will govern the use of smaller versus larger grains. Figure 8 shows the structure of the inner disk of the model for HD 163296, with the grain size regimes illustrated.

The code also allows for the presence of accretional heating of the star. This includes both a magnetically channeled accretion column, and collisional heating of the stellar photosphere.  These are modeled using the ideas of \citet{cal98}. The accretion column produces X-rays, which for simplicity is assumed to radiate uniformly at 100-500 \AA \ over an area a fraction of that of the star,  while the heated atmosphere is emitted as a blackbody uniformly over the star.\footnote{described in the code updates: http://gemelli.colorado.edu/$\sim$bwhitney/codes/codes.html}

Many of the model input parameters use for these two sources are restricted using imaging information. The outer radius of the disk was determined using coronagraphic imaging obtained using the \textit{Hubble Space Telescope} \citep{gra00,ste07}. In the case of HD 163296, the scattered light images reveal a radial surface brightness distribution that is somewhat radially-dependent and azimuthally-dependent, but close to an $R^{-3}$ power law. This requires a disk scale heigh parameter $B \sim1.0$. For our model, $B = 0.99$, and predicts the correct surface brightness distribution at radial distances sampled by the imaging. The disk of HD 31648 was undetected in the survey of \citet{gra00}, but later detected by \citet{ste07} and found to have a steeper radial dependence consistent with greater dust settling. For HD 31648, our model uses $B = 0.6$, an ``anti-flared'' disk. In the case of HD 31648 the predicted surface brightness is not as steep as the observed one, and may be affected by the presence of grain scattering phase functions that are different than those of the model grains. Because grain growth and coagulation are likely to lead to porous non-spherical grains, investigating the properties of those grains in the context of the radiative transfer in disks is warranted.

Similarly, the disk inclination for HD 163296 was set to be 50$^{\circ}$, within the range (45$^ {\circ}$ and 65$^ {\circ}$) reported in the literature \citep{ms97,gra00,was06,ise07}. For HD 31648, the 30$^{\circ}$ inclination is close to the 38$^{\circ}$ of \citet{sim01} and 31$^{\circ}$ of \citet{pi06} (both with quoted uncertainties of $\pm1^{\circ}$).

In the models for HD 31648 and HD 163296, the model outflow regions are empty of material (we are not attempting to model the jet material), but a small envelope has been included, with an  outer radius that is much smaller than that of the disk. In these models, because these two stars are ``isolated'' from their parent star-forming clouds, we do not expect a true ``infalling envelope'' of material that this component of the model is usually used to describe. Here, we are simply using it to simulate the presence of material outside the disk in the inner few AU, where the variability activity is occurring. The inner disk edge is directly illuminated by stellar light and therefore creates excess flux in the 1-5 $\mu$m region, which is seen in the SEDs of both stars (albeit to a much smaller degree in HD 31648 than in HD 163296). The small envelope provides some of the flux in the 8-12 $\mu$m region. The emission from this envelope also reheats the portions of the disk farther from the star, creating an excess of disk flux over that from stellar emission alone in the mid and far IR regions. This component was necessary for obtaining adequate fits in the near-mid IR, and its existence, at least in the case of HD 163296, is also indicated  by the extended  ``halo'' component detected by the interferometry observations of \citet{mon06}. However, in all of the models, the disk surface dominates the flux in the 3 -13 $\mu$m region. In the case of HD 163296, for example, 90\% of the flux is from the disk, and 10\% from the envelope. For the disk flux, less than 1\% is from accretional heating. The rest is from direct irradiation from the star, supplemented by additional warming from the envelope. Figure 9 illustrates the temperature and density structure for the HD 163296 model.

It should be noted that in these models the disk masses are higher than those usually derived from millimeter data. For example, \citet{mks97} provide two estimates of the disk mass for HD 31648. The first assumes an optically thin source at 1.3 mm, a radially-averaged temperature of 40 K, and a grain opacity $\kappa_{\nu}$ of 0.1($\nu$(GHz)/1,200)$^{\beta}$ cm$^{2}$ g$^{-1}$. This yields a mass of gas and dust of $M \approx F_{\nu}d^{2}/\kappa_{\nu}B_{\nu}(T)$ = 0.024M$_{\sun}$ for a gas/dust ratio of 100 and $\beta$=1. A second estimate using power law radial temperature and density gradients results in a disk mass about twice as large. These estimates depend on the assumed dust opacity law $\kappa_{\nu}$ (among other things). In the case of HD 31648 the grain file used for our models has a 1.3 mm opacity only 3\% that used by Mannings et al., which results in a substantially more massive disk. Use of the ``large'' grain population for the disk mid plane (as in HD 163296) would lower the disk mass slightly, but not substantially, since the emission is dominated by the warm disk surface in these models. Extending a large grain model to the surface would result in a 10 $\mu$m band much weaker than that observed, unless the ``envelope'' were made more massive.

In this paper we investigate two possible sources of NIR variability in these systems. These are similar to those discussed by  \citet{dejan07}  for low luminosity systems, and are illustrated in their Figure 12. In their model for low-luminosity systems, they postulate both (A) a puffed-up inner disk and (B) the ablating of a portion of the dust in the inner disk region into a halo above the disk. We will show that changes in one or both may be the source of the variations seen.

\subsubsection{Changing Dust Sublimation Zone (DSZ)}

In the case of HD 31648, we are dealing with an SED that is well into the Meeus Group II category, and generally consistent with the faintness of the outer disk in scattered light. To meet these restrictions we used a disk that was essentially anti-flared, its surface curvature defined by a surface scale height \textit{h} which varies a radial distance \textit{r} as $H \propto R^{0.6}$  (slim disk) and extends from 0.2 [smaller than that measured interferometrically by \citet{ise06}] to 250 AU. For acceptable fits to the size of the 10-$\mu$m silicate band and overall SED, the entire disk utilized the grain model of \citet{cot01}, as using the larger grains of \citet{wood02} led to a poorer fit at the longest wavelengths. The envelope above and below the disk extends from 0.2 to 10 AU and uses the ISM-like grains of KMH. The input star temperature was 8250 K. Accretional heating of the star was included. This heated stellar photosphere produces much of the flux in the 0.1-0.2 $\mu$m region, and emission of this nature is required for matching the UV flux in this star \citep{blo06}. For HD 31648, the ``coverage'' of the X-ray accretion column was 4\%.

Because the details of the fit of the inner region of the disk will depend on the data set used, in Figure 10 we show the model for the 1996 BASS observations only. In Figure 11 we show both the 1996 and 2004 BASS observations, along with their respective models. The model for the epoch 2004 data is essentially the same as that for 1996, but increasing the scale height of the disk at its inner edge from 0.0076 AU to 0.013 AU, and a slight increase in material in the envelope. These changes were introduce solely to be able to approximate the change in the SED. Changes in the inner structure of the disks in some PMS stars has been detected in recent interferometric observations (Tannirkulam, private communication), but simultaneous interferometry/SED observations will be needed for confirming the precise nature of these changes.  It is currently unclear if there was a significant change in the location of the inner radius (through increased grain sublimation), because the weakness of the NIR hump compared to the star'Õs photospheric emission makes constraining the radius of the inner disk difficult using SED information alone. 

In the current version of the radiative transfer code, the disk scale height is set by the requirement that the disk be in hydrostatic equilibrium throughout the disk, and setting the density scale height at the inner edge determines the scale height throughout the entire disk since it uses a single scale height power law exponent. In reality, the sound speed is insufficient for the entire disk to respond so quickly, and the system cannot be characterized by a single scale height power law.  Models that include such dynamic processes are beyond the scope of this investigation.

It is also unrealistic to attempt to fit the entire SED in complete detail in any case, as the lack of simultaneous observational coverage at all wavelengths would make such comparisons of limited value, except to derive disk characteristics in the most general way. In principle, an increase in scale height of the inner disk wall should alter the degree of shadowing of the outer disk, and hence its emission. Obtaining complete simultaneous 1-100 $\mu$m SED coverage will be difficult. Coordinated 1-13 $\mu$m spectral coverage coupled with coronagraphic observations of the scattered light in the outer disk, however, may be a viable alternative to test for the presence of variable shadowing.

For HD 163296, the disk is marginally anti-flared, its surface curvature defined by  $H \propto R^{0.99}$, and the disk extends from roughly 0.3 AU [compared to 0.45 AU derived from the interferometry of \citet{mon06}] to 450 AU. The envelope extends from 0.9 to 10 AU and consists of KMH grains. The disk begins with KMH grains near its surface, but  transitions to the large grains of Wood et al. for the midplane [use of the smaller grains of \citet{cot01} alters the emission only slightly at sub-millimeter wavelengths]. For the model fitting, we used the ``quiescent'' state BASS spectrum from 2005.  The input star temperature was 8750 K, consistent with the spectral type of the star. The accretional heating near the star is less noticeable in this star, due to its earlier spectral type and subsequently greater intrinsic photospheric UV flux. This model is shown in Figure 12. The 2002 ``outburst'' can be fit with the same model, but with an increase in disk height from 0.012 AU for the disk state in 2005 to 0.018 AU for the earlier 2002 state (Figure 13).   In the case of HD 163296, the greater strength of the 1-5 $\mu$m emission makes the location of the inner dust sublimation zone somewhat easier to constrain than in HD 31648, and was different in the two epochs, being roughly 0.29 AU in the low flux stare and 0.35 in the higher state. The fact that these values are some what smaller than those of \citet{mon06} probably indicate that part of the shortest-wavelength emission of the 3 $\mu$m hump is contaminated by emission from hot gas interior to the DSZ. This gas has been recently resolved using interferometers \citep{tat07,ajay07}.

\subsubsection{Detachment of Dusty Material in a Halo}

In the previous section we assumed that a change in the location of the DSZ produced the observed variation in the SED.  However, such variations might also result from dust that is removed from the inner wall region and driven outward, as suggested by \citet{dejan07} in their component (B). In Figure 4 we show the difference in the flux levels in HD 163296 between 2002 and 2005, and model the expected thermal emission of grains capable of producing such a difference, using a simplistic ``model'' based on standard Mie calculations for the expected emissivity \citep{bh83}.  The net emissivity will depend on the size distribution of the grains and their mineralogical content. For our model we have used the optical constants of \citet{hen99} for graphite and pyroxene silicates, $(Mg_{x}Fe_{1-x})SiO_{3}$ with $x=0.8$. Because the disk has probably entered the phase where grain growth and sticking may have proceeded onto the cometesimal-forming phase,  we  adopted a particle size distribution consistent with those measured {\it in situ} for Comet 1P/Halley \citep{mcd87,mcd91}. The maximum and minimum size limits (which were assumed for simplicity to be the same for both materials), the silicate/graphite abundance ratio, and the (single) temperature of the grains were adjusted to reproduce as best as possible the general shape of the observed difference. The final grain size ranged from 0.03 $\mu$m to 25 $\mu$m, with a silicate/graphite ratio by number of 0.5 and a grain temperature of 800 K. Our goal is not to create a definitive description of the ``outburst'' component of the SED, but simply to illustrate that the major features are consistent with the presence of an added population of small grains illuminated by the star. Such grains, when driven away from the star by either winds or radiation pressure, may account for the ``halo'' components suggested by \citet{mon06} for some PMS disk systems,  which in the specific case of HD 163296 was 5.0\% $\pm$  2.5\% at $\lambda$=1.65 $\mu$m. Also shown in Figure 4 is a similar model for the difference in flux for HD 31648 between 1996 and 2004, based solely on the 3-13 $\mu$m BASS data.

For comparison with the model for HD 163296 described in the previous section, both models are shown close up in Figure 13, while in Figure 14 we show the entire SED for 2002, with the ``outburst'' grains providing the extra emission in 2002.

\section{Discussion}

\subsection{HD 163296}

The 1-5 $\mu$m spectral region that underwent the outburst in 2002 corresponds to the  ``hump'' that is commonly seen in the SEDs of many HAeBe stars, and which is generally attributed to emission from the innermost ``wall'' of the dust disk \citep{natta01}, where the dust is subliming (the DSZ) and the incident stellar flux hits the dust layer orthogonally \citep{dd04}. If so, the brightening seen in 2002 would require either increased illumination from the star, or an increase in the surface area of the ``wall'' or some equivalent nearby structure. 

A comparison of the VNIRIS spectra in 2002 and 2005 indicate that the photospheric flux at 0.8 $\mu$m was constant, so that the outburst in the 1-5 $\mu$m region cannot be due to variable heating by the the star at visible wavelengths. However, because the accretion shock contributes to the heating of the disk, the possibility of heating by far-ultraviolet and X-ray photons produced in that region cannot be excluded. Otherwise, the variable emission must be due to changes in the disk itself, and in one instance we will discuss later (the T Tauri star DG Tau) only changes near the DSZ can account for the changes observed.

\citet{dejan07}  discussed two ways that such variability might occur in low luminosity PMS disk systems. One is increasing the scale height of the inner edge of the  inner dusty disk, which is puffed up because the low mass accretion rates result in the inner gaseous disk being optically thin in the radial direction. The other is by the removal of dusty material from the disk into a halo above the disk. Both are plausible, since the former is just a time-variable modification of the standard Dullemond \& Dominik model for shadowed disks, which must occur if the shadowing by accreting gas changes, while the latter is consistent with the excess ``halo'' emission suggested by the interferometry observations by Monnier et al. We have shown that both are capable of reproducing the general character of the near-IR variability, and both may be operating simultaneously.

In both models,  however, the presence of small silicate grains in the wall poses some problems for the standard puffed-up inner disk model. Recently, \citet{dejan06} have shown that a wall with a vertical inner edge can only explain the near-IR emission if the dust has perfectly gray opacity, which is contradicted by the presence of the silicate feature at 10 $\mu$m produced in this region, as it seems to be in these two objects.  The characteristic temperature of the dust in the ``outburst'' in HD 163296 places the location of the grains at or very near that of the DSZ, and because the variations we see in HD 163296 occur simultaneously in both the hump and the silicate band, the simplest explanation is that the small grains are located at or near the wall, and that the opacity cannot be gray. As suggested by \citet{mg07}, this re-opens the debate concerning disks vs. envelopes as the source of the near-IR hump and other spectral features. Highly curved inner walls, which allow the incident radiation to arrive at a more glancing angle of incidence, may avoid these difficulties \citep{ise06,mon06}, but until the models implied by the interferometry can reproduce the SED of this region more accurately, it is unclear whether invoking this geometry alone will solve this problem. 

In Figure 4 we apply the HD 163296 grain model to the difference between the 1996 and 2004 flux levels for HD 31648. Here a higher temperature (T = 1600 K) was required to fit the 3-5 $/mu$m SED. In this case, the grain model had a minimum grain size of 0.24 $\mu$m, as suggested by Isella, and a maximum size of 3 $\mu$m.  It clearly underestimates the silicate band strength. If 0.2 $\mu$m were a ``typical'' grain size in this region, it would have to be the only grain size. Such a monodisperse grain population seems unlikely.

Is the NIR variability related to accretion activity? Figure 15 shows the Ca II triplet lines near 0.85 $\mu$m, as well as the hydrogen Paschen $\beta$ and Bracket $\gamma$ lines for HD 163296 obtained with VNIRIS in 2002 and 2005. While these  lines have usually been interpreted as a measure of accretion activity  in PMS stars \citep{muz98,muz01}, the net strength of the hydrogen bands did not change between 2002 and 2005, while that of Ca II were actually \textit{lower} during the ``hump outburst'' of 2002, compared to the more quiescent state in 2005. This would suggest that this measure of accretion activity can vary out of synch with increased disk wall emission. It has recently been suggested that much of the strength of the hydrogen emission lines may arise fairly far out in the gaseous disk \citep{tat07,ajay07}, and be less coupled to changes in accretion activity very close to the star, where we would expect changes to be most rapid.

\subsection{HD 31648}

HD 31648 exhibits large variations in its IR flux over the 1-13 $\mu$m  wavelength range, and its 1-5 $\mu$m hump implies a smaller DSZ scale height than in HD 163296. In the models of \citet{dd04}, such a small hump,  combined with an even weaker 60 $\mu$m flux, and with millimeter fluxes within 3.5 dex of the bump flux  requires that most of the mass (99.9\% or more) be in very large grains (although our model in Figure 10 did not use the ``large'' grains of \citet{wood02} in the disk midplane). The faintness of the scattered light in the outer disk in the coronagraphic observations of \cite{gra08} is consistent  the coagulation of small grains into large ones and a high degree of settling. In such a model the dust settling leads to a disk that is completely shadowed by the inner wall. 

The structure of the inner disk region of HD 31648 has been probed using interferometry at NIR wavelengths. Measurements obtained using the Palomar Testbed Interferometer in the K (2.2 $\mu$m) photometric band  \citep{ise06}, however,  are only roughly consistent with the actual SED of this object. HD 31648 was also recently observed with the IOTA interferometer \citep{mon06} in the H photometric band (1.6 $\mu$m), and the observations fit best using an inner scattering dust ring with rounded edge and diameter 0.50$\pm$0.04 AU. However, this model also fails to reproduce the bulk of the SED photometric data on this object. When comparing the model SED with observed SEDs taken from the literature, Monnier et al. included no error bars in the literature SED data they plot, as the scatter in the SEDs (source variability) dominated over the individual measurement uncertainties of the SEDs themselves. This variability is clearly illustrated in our Figures 3 and 6, and reinforces the necessity of obtaining interferometry \textit{and} photometric SED observations simultaneously. However, as seen in these figures, it is not simply the object's variability that poses potential difficulties for the models. At no time does HD 31648 exhibit a truly strong hump (spectral slope of zero near 2 $\mu$m in units of $\lambda$F$_{\lambda}$) except in its highest flux state (Figure 11). The success or failure of fitting the model SEDs derived from interferometric observations to photometrically-derived ones will depend on having contemporaneous photometric data. Furthermore,  a substantial amount of the emission at 2 $\mu$m emission seems to be coming from gas interior to the DSZ \citep{ajay07} in some of these systems, and this may apply to HD 31648. 

Other potential issues in model SEDs based on NIR interferometry of  PMS disk systems have been pointed out by by \citet{pont07}.  One is that coverage of the u-v plane is not always optimal for the extraction of precise spatial scales. For example, using a number of different types of spatial models (disks, gaussians, rings, etc.), \citet{eis04} found a range of a factor of two in acceptable angular scales for the ``size'' of the inner disk in HD 31648. \citet{ise06} found a range of acceptable values for inner rim, 0.53-0.63 AU, for the same disk. 

Another issue is that of ``grain survival''. Because smaller grains tend to be hotter than larger grains, there is likely a change in grain size with distance from the star, as larger grains can be closer without undergoing sublimation. In many of the models based on interferometry, only a single grain size is used. Isella et al. used a grain radius of 0.2-0.3 $\mu$m for their grain size, which they describe as  ``big'' (i.e. the large end of ISM grain size distribution). Pontoppidan use 0.3 $\mu$m grains for their small grains and 3.0 $\mu$m for their larger ones. They found that the distance to the inner rim of the disk in the Herbig star VV Ser derived by Isella et al. based on NIR  interferometry was about half that derived  from their own SED fitting. Their best-fitting VV Ser model for the SED and interferometry taken together has their big grains at 0.27 AU and small grains at 0.80 AU.  A similar ambiguity may apply to HD 31648. In fact, the geometry used by Pontoppidan et al. shares many of the features of the models used here.

Objects like HD 31648 present challenges to our understanding of disk structure and evolution. The weakness of the scattered light in coronagraphic imaging experiments and the overall shape of the SED would normally indicate significant dust coagulation and settling has occurred, features that are characteristic only of the later post-accretional  evolutionary stages of disk evolution. Yet HD 31648 is still actively accreting and producing polar jets with HH objects. Any shadowing must also be accomplished with an inner dust sublimation wall of very small vertical scale, since the 1-5 $\mu$m hump in the SED is so weak.  In our models it is less than 0.02 AU high. Taken together, these facts suggest that the disk in HD 31648 may be more like a ``classical'' accretion disk with an inner gas disk merging smoothly with the gas+dust zone, with the former providing enough shadowing to prevent the dust sublimation zone from becoming too puffed-up. But such actively accreting disks generally occur in young systems with highly flared disks, which is  not the case here. A disk with the sort of extreme grain growth and settling  that produces an SED like that HD 31648 should no longer be actively accreting gas. Exactly how low the accretion rates can be and still provide adequate shadowing is addressed by \citet{muz04}, where the gas disk becomes radially optically thin in the mid-pane for accretion rates between 10$^{-8}$ and 10$^{-9}$M$_{\sun}$ yr$^{-1}$. In our model the accretion rate is 4x10$^{-9}$ M$_{\sun}$ yr$^{-1}$ \citep{ste07}, right where this transition may occur. Even in the model of Muzerolle et al., though, only a small fraction of the dust wall is actually shadowed. This star is likely evolving into a transition disk object.

\subsection{Disk Instabilities at the Dust Sublimation Zone}

As yet there is no clear explanation for the source of the changes in and around the DSZ that give rise to the variations on the emission observed in these systems. A number of possible sources of such changes are internal instabilities will be discussed in turn.

\subsubsection{``Standard'' $\alpha$-Disk Instabilities}

Instabilities in long-lived disks can occur in a quasi-periodic manner if there is more than one ``stable'' equilibrium configuration that lead to a limit cycle behavior, as occurs in some cataclysmic variables. The most likely location for instabilities to occur is where there are significant changes in the radiative opacity, equation of state, or other important source parameters within the system. One such location is the transition from an optically thick region dominated by the dust opacity and the dust-free inner gaseous disk (which in principle might be optically thick or thin): the inner boundary of the DSZ. While it is beyond the scope of this paper to determine the precise physical mechanism for the behavior of these disks, we can place some useful constraints on the nature of these changes by comparing them to what is expected from different instability time scales. Because the variability is primarily in the region responsible for the NIR hump,  the boundary between where the radiative opacity being controlled by dust to where it is being dominated by gas is implicated. 

Can instability-driven changes in the disk structure in the vicinity of the inner ``wall'' of the DSZ lead to the sort of changes observed?  Because the disk surrounding HD 163296 is still actively accreting, it can be heated by both the radiation from the star and from internal viscosity. At the present time, the stability of disks heated both internally and externally is not well understood, but we can get an approximate idea of variability time scales, some of which do not depend explicitly on the degree of viscous energy dissipation. 

The general physics of such disks are discussed by \citet{ss73}, while \citet{pring81} outlines some of the instabilities and their timescales in these systems. Typical timescales encountered in accretion disks are the dynamic timescale $t_{\phi}$, the hydrostatic timescale $t_{z}$, the  thermal timescate $t_{th}$, and the viscous timescale $t_{\nu}$. The dynamic timescale is the time required for a parcel of mass in the disk (such as a hot spot) to move a significant fraction of its orbit, and  is given by

$t_{\phi}=\frac{R}{v_{\phi}}=\Omega^{-1}$

\noindent where R is the radius outward in the disk, $v_{\phi}$ is the orbital velocity, and $\Omega$ is the angular velocity (note that the orbital period is 2$\pi$$ t_{\phi}$). For R in AU, $t_{\phi}$ in years, and the stellar mass M in M$_{\sun}$, Keplerian orbits will have

$t_{\phi}=\frac{1}{2\pi} \frac{R^{3/2}}{M^{1/2}}$.

\noindent Depending on the disk inclination and thickness, any hot spot might be more easily visible at some time than at others.

The timescale required for a region of scale height H and sound speed $c_{s}$ to establish hydrostatic equilibrium is

$t_{z}=\frac{H}{c_{s}} \sim t_{\phi}$.

\noindent For changes in the 1-5 $\mu$m feature, the relevant distance is the dust sublimation zone, which for blackbody-like grains  will be located approximately at

$R \sim (\frac{278}{T})^{2}(\frac{L}{L_{\sun}})^{1/2}$ AU.

\noindent For a star with $L \sim 26 L_{\sun}$ and a grain evaporation temperature of $T \sim 1500 K$, the inner dust wall would be at $R \sim 0.3 AU$. In this case, 

$t_{\phi} \sim t_{z} \sim$ few days

\noindent for a star whose mass is $\sim 2.5 M_{\sun}$.

By contrast, the thermal time scale is the time required for a disk to dissipate a sizable fraction of its internal heat, and is given by

$t_{th} \sim \frac{1}{\alpha} t_{\phi}$ 

\noindent where $\alpha$ is the standard dimensionless viscosity parameter \citep{ss73}. \citet{lee05} has suggested that, for a variety of reasons, $\alpha \sim 0.001$ is a good reasonable global average for the accretion history in disks surrounding T Tauri stars, with 0.01 being an upper limit. If true for the systems being discussed here,  $\alpha \sim$  0.01-0.001 would give

$t_{th} \sim$ 10-100 years,

\noindent Finally, the viscous time scale, over which the disk surface density can change, is 

$t_{\nu} \sim (\frac{R}{H})^{2} t_{th}$.

\noindent For a thin disk $t_{\nu}$ must be considerably longer (e.g., thousands of years).

It is striking that the IR variations at the disk wall have time scales comparable to that of the ejection of HH objects. However, the disk instabilities near the location of the dust sublimation zone cannot induce accretion events on time scales shorter than the length of time required to accrete matter from the wall to the star, and this is also measured in thousands of years or longer. 

While it seems unlikely that changes in the region of the DSZ could directly impact the accretional activity on time scales shorter than centuries, changes in the inner gaseous disk will have much smaller time scales and could impact both the DSZ and the accretion rate (an subsequent gas outflow in the jets). In the case of HD 163296, a ``rapid'' (i.e., time scales of weeks to months) change in the scale height of the inner gas disk could lead directly to changes in the illumination of the dust wall. This will create a widening zone of subliming grains whose outer radius will move outward from the star until the grains are cool enough to survive. The changes observed in the dust scale height (whose rapidity will be on the order of  $t_{z}$) and inner radius (governed by how rapidly the grains can sublime) are consistent with such a scenario.

\subsubsection{Magneto-Rotational Instabilities}

If the inner dusty disk is partially ionized, then a magneto-rotational instability \citep{bh91a,bh91b} could also develop. Such instabilities have been recently invoked to describe gas accretion without equally efficient dust accretion  in actively-accreting objects with ``cleared'' inner disk holes \citep{chiang07,kretke07}. Such an application is particularly relevant to disks where the main inner disk wall is far beyond the location of the DSZ for \textit{refractory} grains, but may coincide with that of \textit{volatile} grains (the ``snow line''). Here the instability is invoked to help explain the presence of accretion in what would otherwise be non-accreting systems. The main requirements are ionizing radiation from the star - usually assumed to be X-rays from both the accretion columns and stellar corona with magnetic fields embedded in the ionized gas. Both HD 31648 and HD 163296 possess hot accreting gas (and HD 163296 is an X-ray source), and the highly collimated outflows suggest significant magnetic fields in the star-disk region, providing the main ingredients for MRI processes. In the models presented here, the X-rays and ultraviolet emission from the accretion shock are responsible for some of the heating of the disk. Changes in accretion onto the star must alter the heating of the exposed dust wall. While the MRI model has generally been invoked to provide a mechanism for \textit{steady} accretion, the details of how time-dependent changes would occur, and their possible effect on the DSZ, are not known. But such a mechanism might provide a direct link between activity near the surface of the star (perhaps related to jet activity and HH object production) and changes in the DSZ.

\subsubsection{The X-Wind}

Both accretional heating from within the disk and high energy stellar and accretion shock photons photons hitting the surface can cause the gaseous disk to become ionized. In the X-Wind models of \citet{shu94} (and see the comments in \citet{shu07}) the stellar magnetic field is connected to that of the differentially-rotation gas disk. The model not only provides a mechanism for the production and collimation of the outflowing jets observed in the stars discussed here, but opens the possibility that reconnection events in the magnetic field could alter both the outflow rate and structure in the ionized gas disk at the DSZ interface. Should the X-ray and ultraviolet flux from the accretion shock ionize the surface of the DSZ, then it may also be affected.

\subsubsection{Planetary Perturbations}

Finally, we are observing these stars during a phase of disk development when planets could be forming. Planet formation has been suggested for the truncation of the thick dusty disks far beyond the DSZ for refractory grains in the disk of TW Hya \citep{cal02}, and the thinning of material interior to it, effectively creating a class of ``transition'' disk system beyond the Meeus Group I and II classification. If the structure of the inner wall is in any way affected by such a body having an orbit close to the current sublimation distance (about 0.24 AU), then its orbital period would be on the order of months. So far, we have no indication of changes on time scales that small in HD 163296, and this time scale is an order of magnitude shorter than that of the ejection of the HH objects.

\subsection{Other Observable Consequences}

If changes are occurring in the location and height of the DSZ, these should be observable in a number of different ways.  If some of the stellar radiation intercepts the outer flared disk, then changes in scale height of the inner rim will alter the degree of shadowing of the outer disk. This will produce changes in both the scattered light and the far-IR emission. Multi-epoch observations of the scattered light coronagraphic images of HD 163296 suggest such changes might have been detected \citep{wis07}. Far-IR monitoring with sufficient time coverage and precision does not exist for this star, but such variations may have been seen in the thermal emission of the HAeBe star SV Cep \citep{ju07}.

If the DSZ undergoes changes in exposure to stellar radiation, there would be two other important consequences that could be tested. As exposure increases, newly exposed dust will find itself above the sublimation temperature, and the DSZ must retreat from the star (Figure 16).  When and if the gas opacity and/or scale height is re-established, some of the vaporized material, if not driven away by stellar or disk winds,  may re-condense, and the DSZ would gradually fill back in. This might be observable with NIR interferometers, but would require regular monitoring programs. 

Because the emissivity of grains is selectively poorer for smaller grains than larger ones, the smaller grains would be the first to vaporize. Any surviving grains would be large and would undergo annealing if their temperatures were greater than about 800 K \citep{sh98,sh00,hill01}. Once irradiation of the disk returns to lower values, grain condensation, occurring over time spans of months, could add additional crystalline material to this zone. Ablation of the dust in this region by stellar winds and photon pressure would gradually redistribute some of this material further out, but would still be preferentially be concentrated at its source, the DSZ  region. Because the ratio of grain mass to surface are increases with increasing grain size (at least for compact grains), these processes will transport the smaller grains more effectively, leaving the larger grains closer to the DSZ. Infrared spectroscopy of the silicate band emission using the Very Large Telescope Interferometer \citep{vB04} indicates that the inner few AU of these stars are in fact characterized by both larger grains and more crystalline material, relative to the outer regions, as would be expected with this scenario. Multi-epoch spectroscopic monitoring of a few of these sources with the VLTI would be useful investigating this possible mechanism.

\subsection{One Size Does Not Fit All: The Remarkable Case of DG Tau}

While migration of the DSZ may be sufficient to explain the variations seen in the SEDs of HD 31648 and HD 163296, this mechanism cannot explain the behavior of  other PMS objects exhibiting near-IR variability.

\citet{woo00} first reported unusual changes in the 8-13 $\mu$m spectrum of the T Tauri star DG Tau. In 1996, its IR spectrum consisted of a very weak silicate emission band, with spectral structure consistent with the presence of both amorphous and crystalline silicates. A year later, the emission band was weaker (almost nonexistent), and the amorphous component was undetectable. In early 2001, a dramatic change was captured by two different instruments:  DG Tau exhibited an an overall increase in 3-5 $\mu$m emission, and the onset of a silicate band in absorption \citep{woo04}.  The 2001 BASS spectrum, shown in Figure 17, is (so far) unique for a PMS star in showing the presence of  crystalline silicate material in absorption. Between 2002 and 2004, the star returned to its previous  ``conventional'' behavior. But in 2006 it underwent another 3-5 $\mu$m outburst, but on this occasion, the silicate band was in emission! The breadth of the emission band requires mean grain sizes in excess of a few $\mu$m, smaller than those present in its quiescent state, but larger than the ones responsible for the 2001 absorption event.

The relatively simple DSZ expansion and migration scenario is clearly inapplicable to DG Tau. Rather, a scenario more like the ``ablation'' one discussed by \citet{dejan07} seems more likely. This picture is consistent with the observation by \citet{kit96} of the expansion of the disk in DG Tau, which they surmise may be driven by a stellar wind. However, while the expansion derived from their CO radial velocity measurements are primarily confined to the outer disk, the lofting observed in 2006 requires warm grains, indicating that this radial outflow of disk material may originate near the DSZ in this star.

\section{Conclusions}

We have presented observations of two ``isolated'' Herbig Ae stars that exhibit significant variability in their infrared emission. Our immediate finding is that any model of the SED based on data obtained in a single epoch cannot describe the nature of source completely.  In this sense, {\it there is no ``the model'' for any such object}, and this problem is exacerbated further when modeling includes the use of disparate data sets obtained on different epochs (often using diverse photometric systems).  Our goal is to provide the sort of multiwavelength multiepoch data that models require if they are going to be capable of explaining {\it all} of the data in a self-consistent manner with the fewest changes in model parameters.

HD 31648 and HD 163296 test our ability to understand these disks, because both exhibit features that  are characteristic of PMS disks systems that are both very young and simultaneously relatively old in terms of evolutionary development. Both are sources of polar jets of outflowing material and HH objects which must be supplied by the active accretion of gaseous material close to or onto the star. That is, both also exhibit other indicators of highly active inner regions which are associated with accretion shocks, such as far-UV and X-ray emission and high-temperature ionic gas species. Yet their SEDs in the mid to far infrared have led them to be classified as stars  where the outer disk has undergone significant grain coagulation and settling to the point where much of this region is shadowed by the innermost disk, which has been forced to expand vertically by a paucity of accreting gas.

In both stars, interferometric observations (PTI and IOTA) indicate the presence of an inner disk similar to that expected in models that include a puffed-up inner wall due to direct illumination by the central star. Geometries like these are consistent with the Group II classification and the faintness of the coronagraphic observations of the scattered light in the outer disks. Nevertheless, the sum of the data present us with a number of challenges. 

For HD 31648, \citet{ise06} have remarked that the predicted SED derived from models of the interferometry measurements is ``only roughly consistent with the photometric values.'' Their models generally predict an increase in brightness (in units of $\lambda$F$_{\lambda}$) with wavelength that is not a good match to the photometrically-determined SED. In their model SEDs, the predicted brightness at 3.5 $\mu$m is consistently too large.  An examination of the dates when the PTI observations were made \citep{eis04} indicate that they were obtained over a time span of a year, during which time HD 31648 may have been varying in brightness at the wavelength that these observations were made. The nearly ``flat''  1-2 $\mu$m spectral shape seen in the models derived from the IOTA observations \citep{mon06} are consistent with the photometric data only if HD 31648 was in a bright state when it was observed. To properly derive the structure of these disks,  nearly simultaneous photometric and interferometric measurements will be required, and the effects of hot accreting gas included. 

HD 163296 presents a different problem. Here, the interferometry data, the disk shadowing, and the presence of a strong 1-5 $\mu$m hump are all consistent with the vertically expanded inner wall at the DSZ. But how can the disk be illuminated directly by the star as required by such {\it passive} disk models, yet still be \textit{actively} accreting gas at a rate large enough to drive the bipolar gas flows and HH objects? This requires that the equatorial optical depth of the inner gas disk be less than unity, or its vertical height be much smaller than the height of the dust sublimation wall. This apparent dichotomy may be solved if the inner gas disk is much thinner than the disk wall, and the accretion rate small, but sufficient to shadow at least part of the dust disk mid-plane, as in the model of \citet{muz04}. We may be observing many of these sources just prior to them reaching the transition disk phase, where the accretion of gas is diminishing, but not yet terminated, and may be sporadic. Should the scale height of the disk have a vertical ``step'' - going from the inner region goverened by gas opacity to the outer one governed by dust opacity -  it might result in an inner dust sublimation zone of complex structure, where the equatorial regions are indeed shadowed by the gas, but the upper regions exposed to unattenuated stellar radiation. 

In both stars we may be witnessing the final phase of active accretion. HD 31648 has one of the lowest detected accretion rates in intermediate-mass PMS stars \citep{gra08} and is also amongst the oldest of these stars to exhibit active accretion. The time scale over which changes in the innermost gas disk and jet activity could be on the order of a few years, while that of the DSZ itself must probable be longer. If so, repeated exposure and shadowing of the DSZ will lead to a region inside the DSZ where grains are repeatedly heated and cooled. It is here that one would expect to be able to anneal amorphous grain material into a crystalline phase if they survive sublimation, or condense them directly into crystalline form if they do not. Repeated heating and cooling dust material in the DSZ would enhance the production of crystalline material there, as has been observed by VLTI spectroscopy.

In the models by \citet{ise06} for the inner disk rim, grain sizes ``larger than $\sim$ 1.2  $\mu$m and possibly much larger'' are required to explain the interferometry and the gross features of the SEDs of the Herbig stars they studied. However, much larger grains would be inconsistent with the grain sizes needed to produce a strong  10-$\mu$m silicate band in the ``outburst'' of HD 163296, if that material is associated with the region of the inner disk wall.

Part of the difficulty with matching the models derived from interferometry measurements with the observed SEDs  may result from intrinsic variability of the emission. Interferometric measurements made at different epochs may yield different results, and such changes have been recently confirmed by Tannirkulam \& Monnier (private communication). Complete u-v plane coverage needs to be obtained quickly for a single epoch, along with simultaneous SED photometry, before such observations can be effectively used to construct realistic models of the inner disk region.  On the positive side, multi-epoch coverage will provide a very powerful tool for studying physics of these regions.

Finally, the case of DG Tau serves as a warning that the actual structure of the inner regions of these systems must be complex and highly variable. Expecting a simple time-independent model to describe this system is doomed to failure. Models that include even the most rudimentary dynamical changes are needed.

We are far from being able to claim that we truly understand the structure of PMS disks. For a better understanding of these systems, it is crucial that repeated observations with NIR interferometers be obtained on a more regular basis over the course of at least a few years, accompanied by nearly-simultaneous photometry.  It is our hope that measurements such as these will provide useful constraints on future modeling of the inner disk regions of these objects.

\acknowledgments

We would like to thank Lee Hartmann, Ajay Tannirkulam, and Casey Lisse for useful discussions, Barb Whitney for advice and suggestions on using the Monte Carlo radiative transfer code, and NASA's \textit{Infrared Telescope Facility} for 10 years of access required for this project. Additional thanks must go to the anonymous referees, whose input helped improve the paper considerably. MLS was supported in part by The Small Research Grant program of the American Astronomical Society, the University Research Council of the University of Cincinnati, NASA Origins of Solar Systems grant NAG5-9475, and NASA Astrophysics Data Program contract NNH05CD30C. This work was partially supported by the Independent Research and Development program at The Aerospace Corporation.

\clearpage
\appendix

\section{Appendix A: Infrared Photometry \& Flux Calibration}

While filter technology and our understanding of atmospheric transmission has gradually led to the evolution of ``standard'' filter sets for near infrared (wavelengths shorter than  5 $\mu$m) measurements \citep{st02,my05}, the situation in the 8-14 $\mu$m telluric transmission window is less standardized. 

A variety of infrared magnitude systems have also been in used during the past 3 decades. In many cases, these derive from the different filter systems being used, so that transformations between systems is required. In general, these may be relatively precisely calibrated [for example those between six different systems described in detail by \citet{glass85}]. Some are based on a system where $\alpha$ Lyr has a magnitude of 0.0 at all wavelengths, while in others it is the color index  of an ``average'' A0 star that is defined as being 0.0, while $\alpha$ Lyr itself may be anywhere between +0.02 and -0.02 mag \citep{glass85} or even brighter. 

Somewhat more puzzling are systematic differences between some filter systems and higher spectral resolution spectrophotometric systems. For example, $\alpha$ CMa is a favorite calibration star because it is bright, and because its infrared spectrum contains relatively weak photospheric lines, making the calibrations insensitive to the exact wavelengths of the detectors. Nevertheless, many photometric systems use magnitudes for this star in the 2-14 $\mu$m region that are roughly 1.50 mag brighter than $\alpha$ Lyr \citep{gehrz74}, while spectrophotometric measurements made with BASS \citep{rm98} and other systems \citep{cohen95} indicate a value closer to 1.35 mag., a 15\% difference. 

When actually comparing real absolute flux values (instead of magnitudes), as is usual for SEDs, the choice of an absolute flux calibration system is important. These generally fall into three broad classes: those based on the absolute flux of the Sun, those based on stellar atmosphere models, and those based on terrestrial blackbody sources. The precise system chosen will lead to differences in the derived fluxes. For example, the absolute flux of $\alpha$ Lyr at 8.7 $\mu$m on the BASS system in units of $\lambda F_{\lambda}$  is 1.90x10$^{-11}$ W m$^{2}$, while that of \citet{gehrz05} is 1.8x10$^{-11}$ W m$^{2}$, and that of \citet{cohen92} is 1.71x10$^{-11}$ W m$^{2}$, a range of 10\%. A comparison of the BASS and Cohen systems, along with that used by \citet{sit81} for the first multiwavelength SED study of the stars in this paper, are shown in Figure 16. 

New work has recently been shed light on the application of stellar photospheric models to $\alpha$ Lyr. It has been known for some time that $\alpha$ Lyr sits above the main sequence, and that the width of its spectral lines implied a small value of the projected rotational velocity {\it v sin i}. Optical interferometry by \citet{pet06} support the notion that the star is actually flattened due to rapid rotation (93\% breakup speed)  and seen nearly pole-on. They show that their model predicts excess emission over the standard spherical stellar atmosphere models by an amount that grows from 1.6\% at 0.67 $\mu$m  to 7.2\% at 3.69 $\mu$m. 

Because of all of the effects described above, the blind use of different relative and absolute flux systems can introduce spurious flux differences of up to 17\%.

Despite these differences, for the purpose of studying the variability of SEDs, the precise system to be used is not as important as using (as much as possible) the same system for all observations. For the purpose of this study, we have used the measurements of \citet{rm98} to define the relative brightness of our calibration stars in all systems, and have used the BASS absolute flux calibration throughout. We have re-examined the data reductions used for the observations of the HAeBes reported previously by \citet{sit81} and \citet{sit94}, with the aid of the original observing logs, in order to make a better estimation of the true errors and eliminate possible systematic effects due to changes in seeing, sky transparency, and calibrations stars used, as indicated in those logs. The dates of the observations, telescopes, and instruments used for the infrared observations used here are listed in Table 1.

In addition to the infrared observations listed in Table 1, we have also used the Low Resolution SpectroPhotometer (LRSP) observations obtained with the 0.4-m reflector of the Pine Bluff Observatory of the University of Wisconsin. These data are the same ones used by \citet{sit81} for determining the first mutliwavelength SEDs of these two stars. The instrument utilized a movable grating to sample the spectrum  between 3390 and 5840 \AA, using a photomultiplier tube and photon-counting system. The spectral resolution was set by an aperture mask that could sample the spectrum in bandpasses of 40 and 400 \AA. For the observations presented here, the spectrum was sampled at 17 wavelengths using the 40 \AA \  bandpass mask. The main use of these observations is to tie the infrared data to UV observations obtained with the \textit{International Ultraviolet Explorer}. They have a distinct advantage over filter photometry since they are capable of providing a ``clean'' Balmer discontinuity not possible with the standard broad-band U filter. These data also usually include the peak flux of the stars, which are important for subsequent modeling calculations. They also allow a comparison with filter photometry from the literature, in order to detect possible variability at visible wavelengths.

\begin{figure}
 \center
 \includegraphics[scale=0.9]{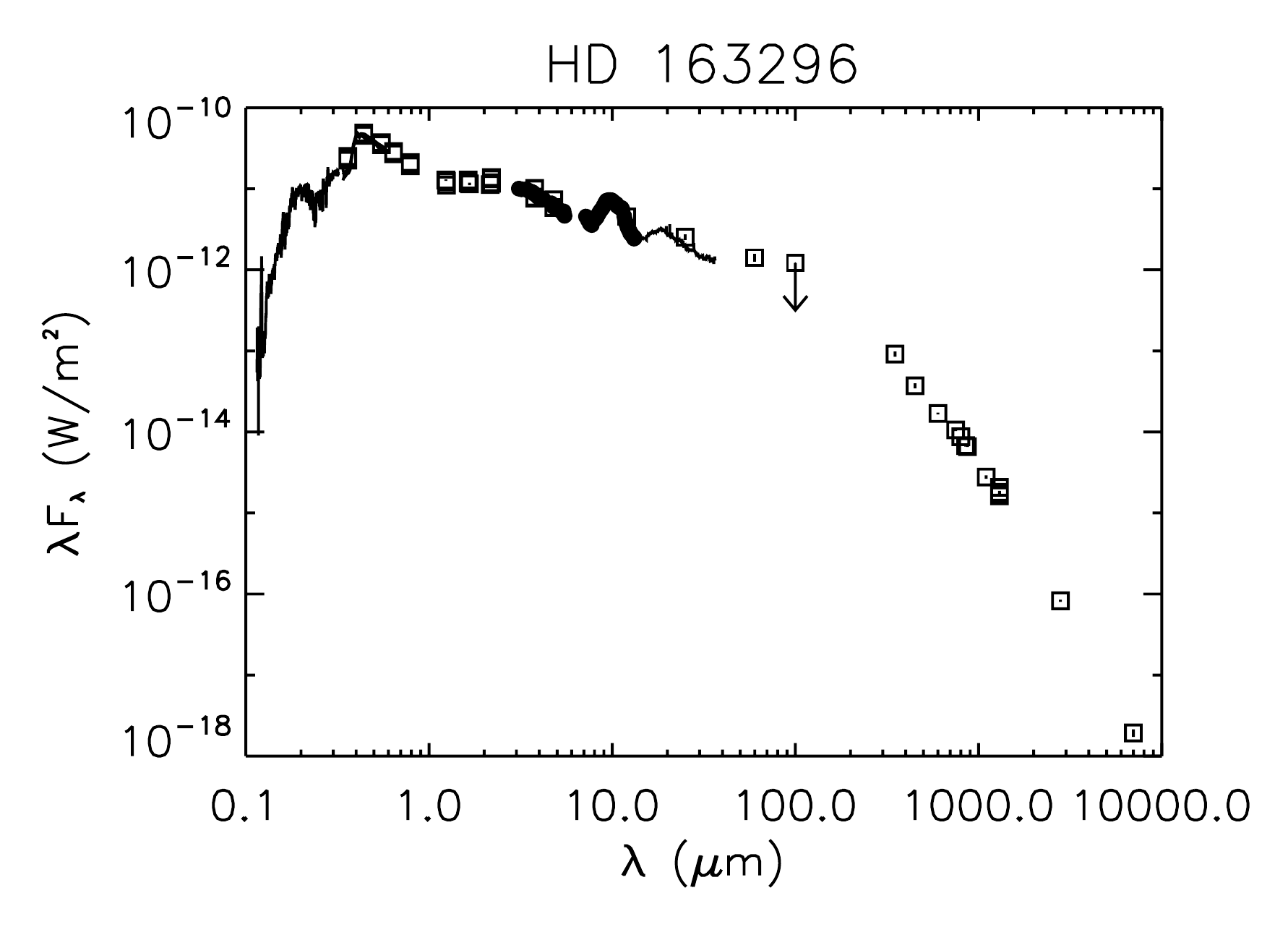}
\caption{The spectral energy distribution of HD 163296. From shortest wavelengths to longest: solid lines -  IUE merged spectrum (LWR05626 \& SWP06566 on 1979 Sep 19) and  Pine Bluff LRSP (1979 Jun 22, 1979 Aug 12, \& 1979 Sep 17);  open squares between 0.36 $\mu$m and 4 $\mu$m - UBVRI photometry [15 sets over  9 nights, from \citet{dolf01}],  JHKLM photometry [4 nights between 1980 \& 1986, from \citet{dolf01}], and 2MASS; solid line between 13 and 40 $\mu$m - Spitzer IRS (2004 Aug 28); filled circles - BASS (1996 Oct 14); open squares from 12 to 100 $\mu$m - IRAS photometry; open squares longward of 100 $\mu$m - millimeter \& submillimeter photometry from \citet{hen94},  \citet{hen98},  \citet{me94}, and \citet{ise07}. \label{fig1}}
\end{figure}
 
\clearpage

\begin{figure}
 \center
 \includegraphics[scale=0.9]{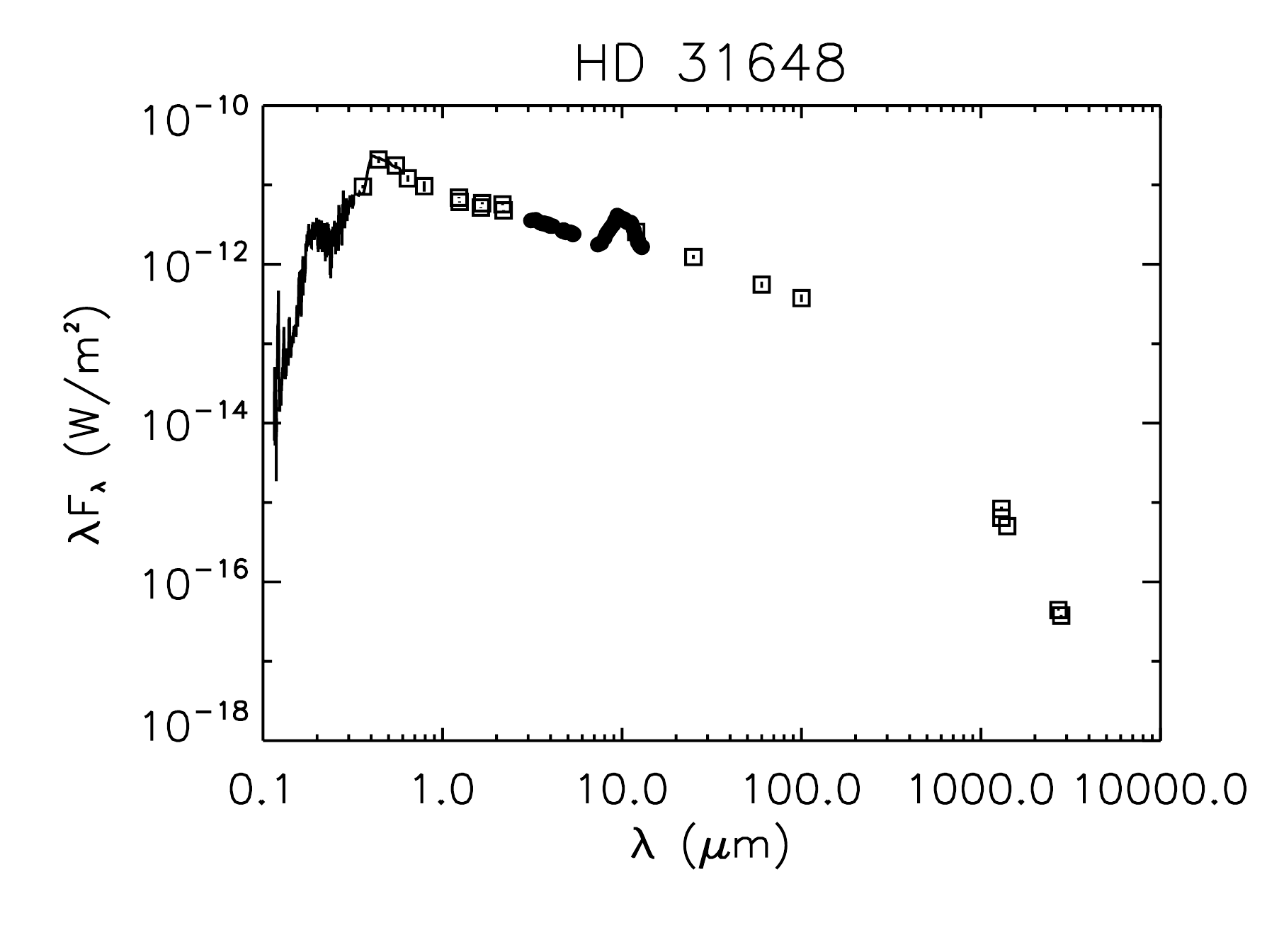}
\caption{The spectral energy distribution of HD 31648. From shortest wavelengths to longest: solid lines -  IUE merged spectrum (LWR04278 \& SWP04940 on 1979 Apr 15) and  Pine Bluff LRSP (1979 Mar 10);  open squares between 0.36 $\mu$m and 4 $\mu$m - UBVRI photometry [1998 Oct 25 from \citet{oud01}],  JHK photometry [1998 Oct 28, from \citet{eir01}], and 2MASS; filled circles - BASS (1996 Oct 14); open squares from 12 to 100 $\mu$m - IRAS photometry; open squares longward of 100 $\mu$m - millimeter \& submillimeter photometry from \citet{mks97}, \citet{ms97}, and \citet{pi06}.  \label{fig2}}
\end{figure}

\clearpage

\begin{figure}
 \center
 \includegraphics[scale=0.9]{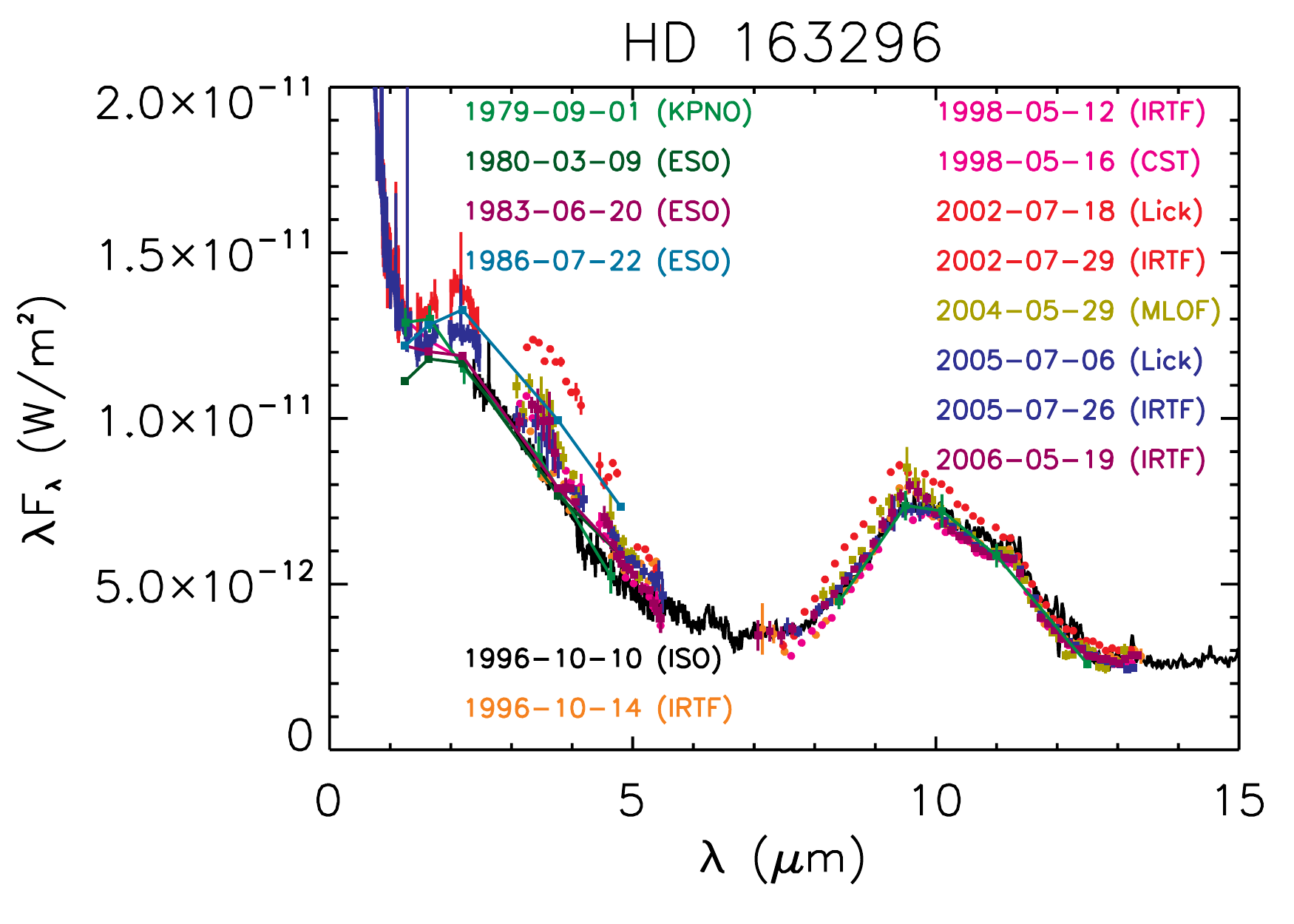}
\caption{The infrared flux of HD 163296 on 14 dates, organized as 10 separate epochs. Due to the proximity in time and similarity in flux level, the following can be considered single epochs: 1996 Oct 10 (ISO) \& 14 (IRTF), 1998 May 12 (IRTF) \& 16 (CST), 2002 Jul 18 (Lick) \& 29 (IRTF), 2005 Jul 06 (Lick) \& 26 (IRTF).  The ESO data are from  \citet{dolf01}; the CST data are from \citet{eir01}. The KPNO data were previously published in \citet{sit81}. \label{fig3}}
\end{figure}
 
\clearpage

\begin{figure}
 \center
 \includegraphics[scale=0.9]{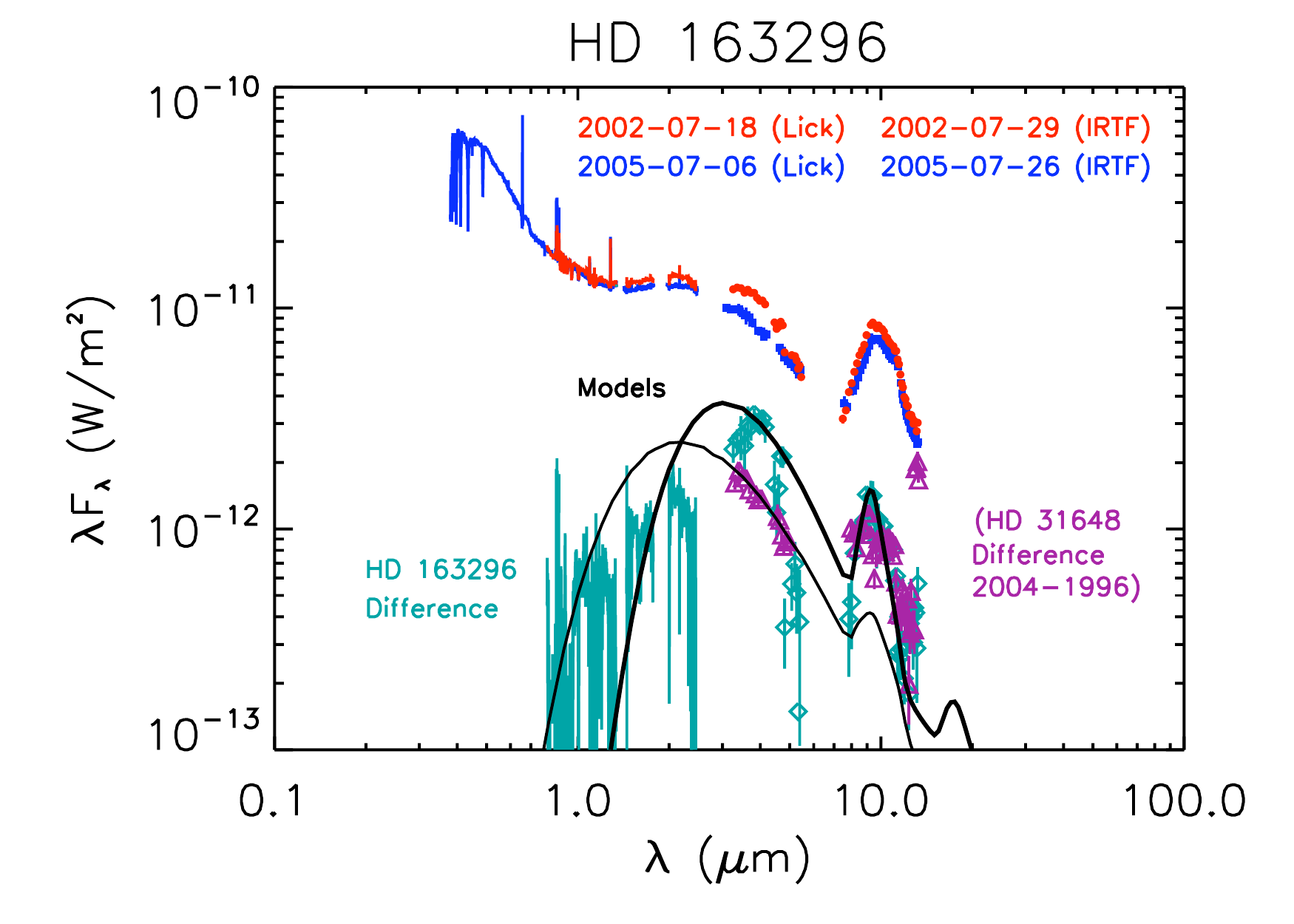}
\caption{The combined VNIRIS+BASS spectroscopy of HD 163296 for 2002 and 2005. The lowest curve and diamond symbols are the difference between the flux levels measured by VNIRIS and BASS, respectively, for the two epochs. The thicker smooth solid line is a model spectrum of a mixture of graphite and silicate grains (the latter being Mg$_{0.80}$Fe$_{0.20}$SiO$_{4}$). The particle size distribution used was that of Dust Impact Detection System (DIDSY) {\it in situ} measurements of Comet 1P/Halley \citep{mcd87,mcd91}, with size cutoffs at 0.03 and 0.25 $\mu$m.  Also shown in the figure is the difference in the 3-13 $\mu$m flux of HD 31648 between 1996 and 2004 (triangles).  The thinner smooth solid line is the same grain model as for HD 163296, but with a temperature of 1600 K, a minimum grain size of 0.25 $\mu$m [consistent with the model of \citet{ise06}] and a maximum size of 3.0 $\mu$m. Such a large minimum size cannot explain the changes in flux for HD 31648. Unless the grain size range is unrealistically restricted,   a ``large'' grain population cannot account for the relatively strong presence of the silicate band in the differenced spectra of these two systems. \label{fig4}}
\end{figure}

\clearpage

\begin{figure}
 \center
 \includegraphics[scale=0.9]{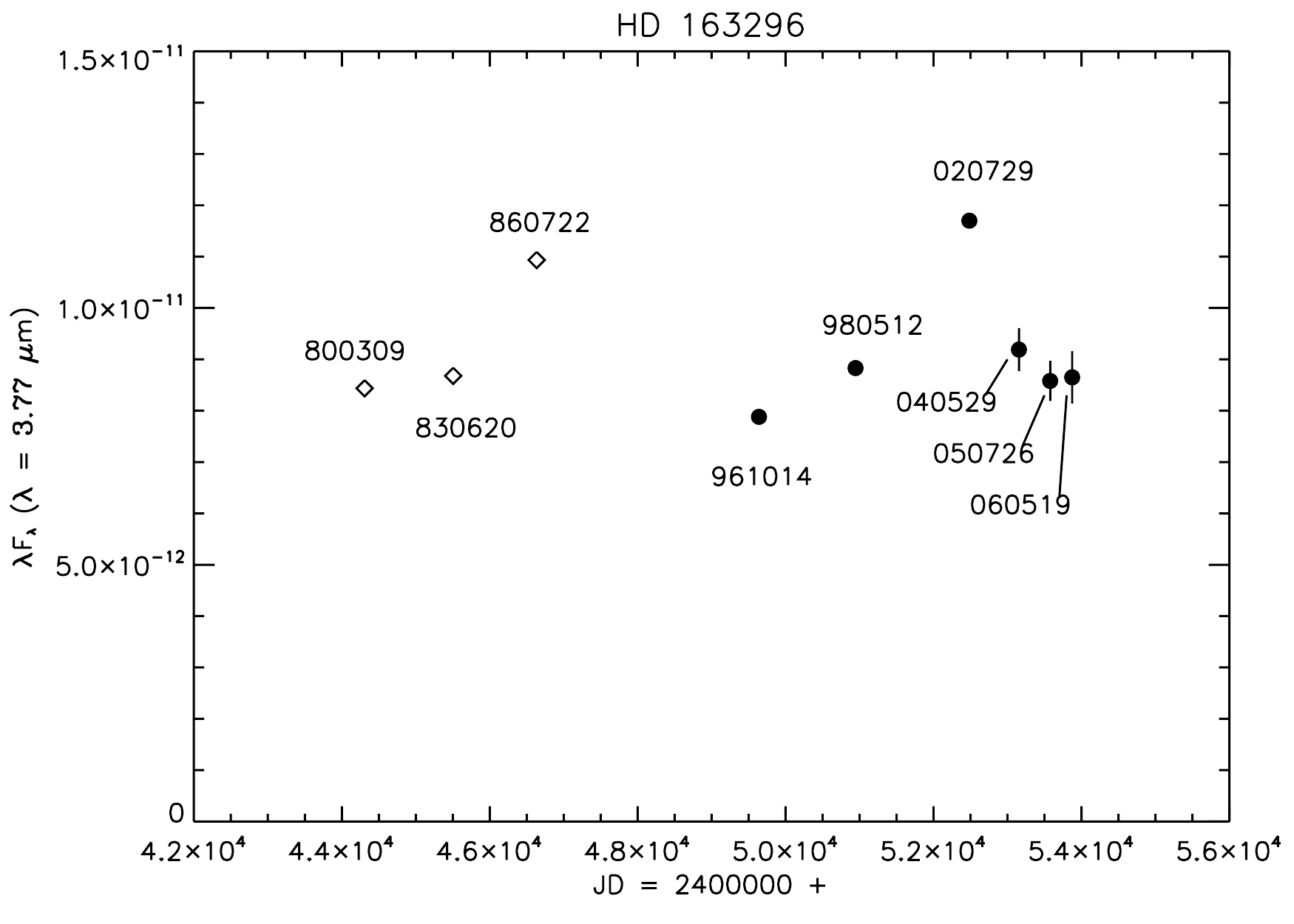}
\caption{Light curve of HD 163296 at 3.77 $\mu$m.  The diamonds are ESO photometry, while the filled circles are derived from BASS/IRTF spectrophotometry. The UT dates of the observations are given as  \textit{yymmdd}. \label{fig5}}
\end{figure}

\clearpage

\begin{figure}
 \center
 \includegraphics[scale=0.9]{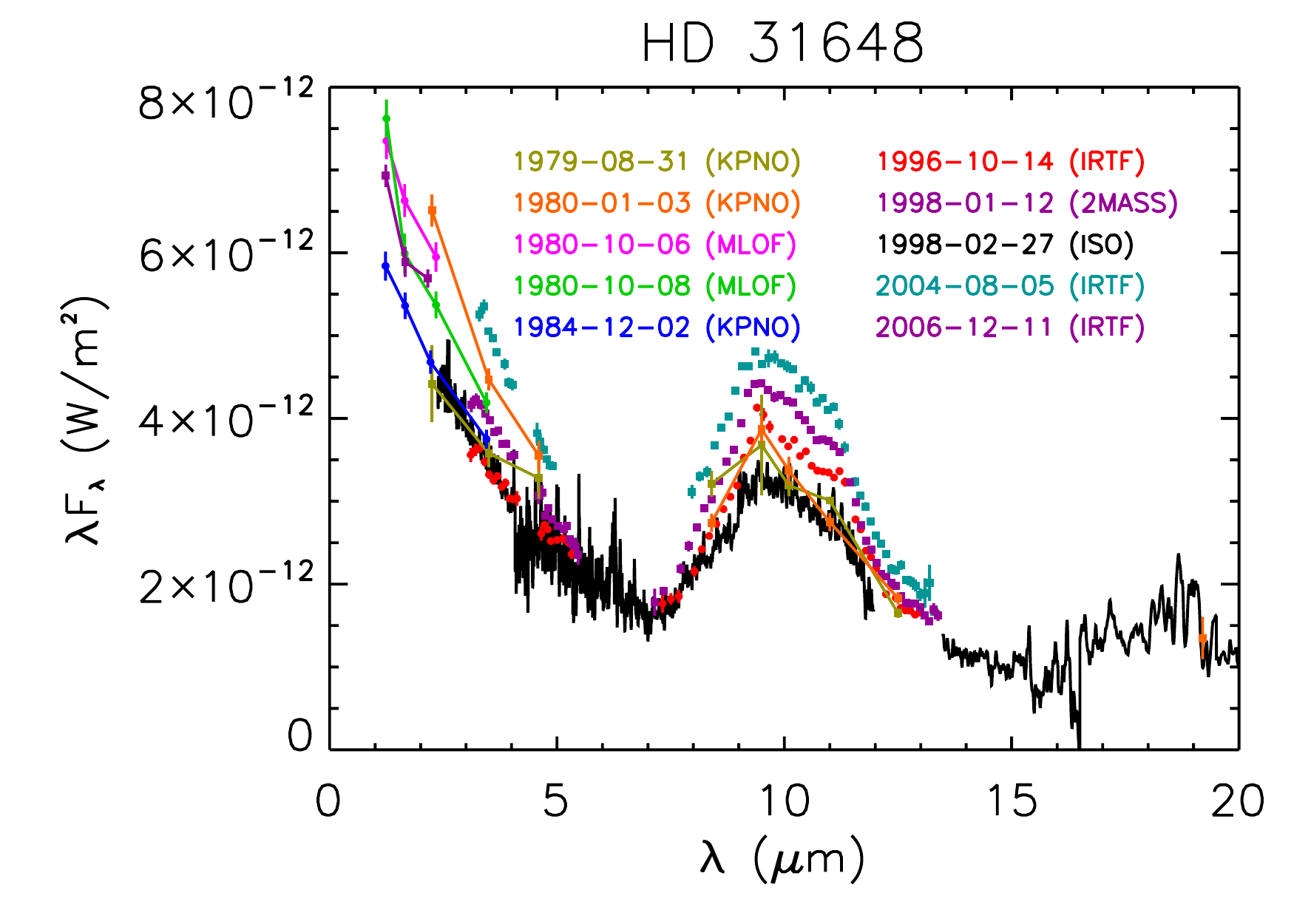}
\caption{The infrared flux of HD 31648 on 10 dates.  The 1979 KPNO data were previously published in \citet{sit81}. The KPNO and MLOF data between 1980 and 1984 are previously unpublished. \label{fig6}}
\end{figure}

\clearpage

\clearpage

\begin{figure}
 \center
 \includegraphics[scale=0.9]{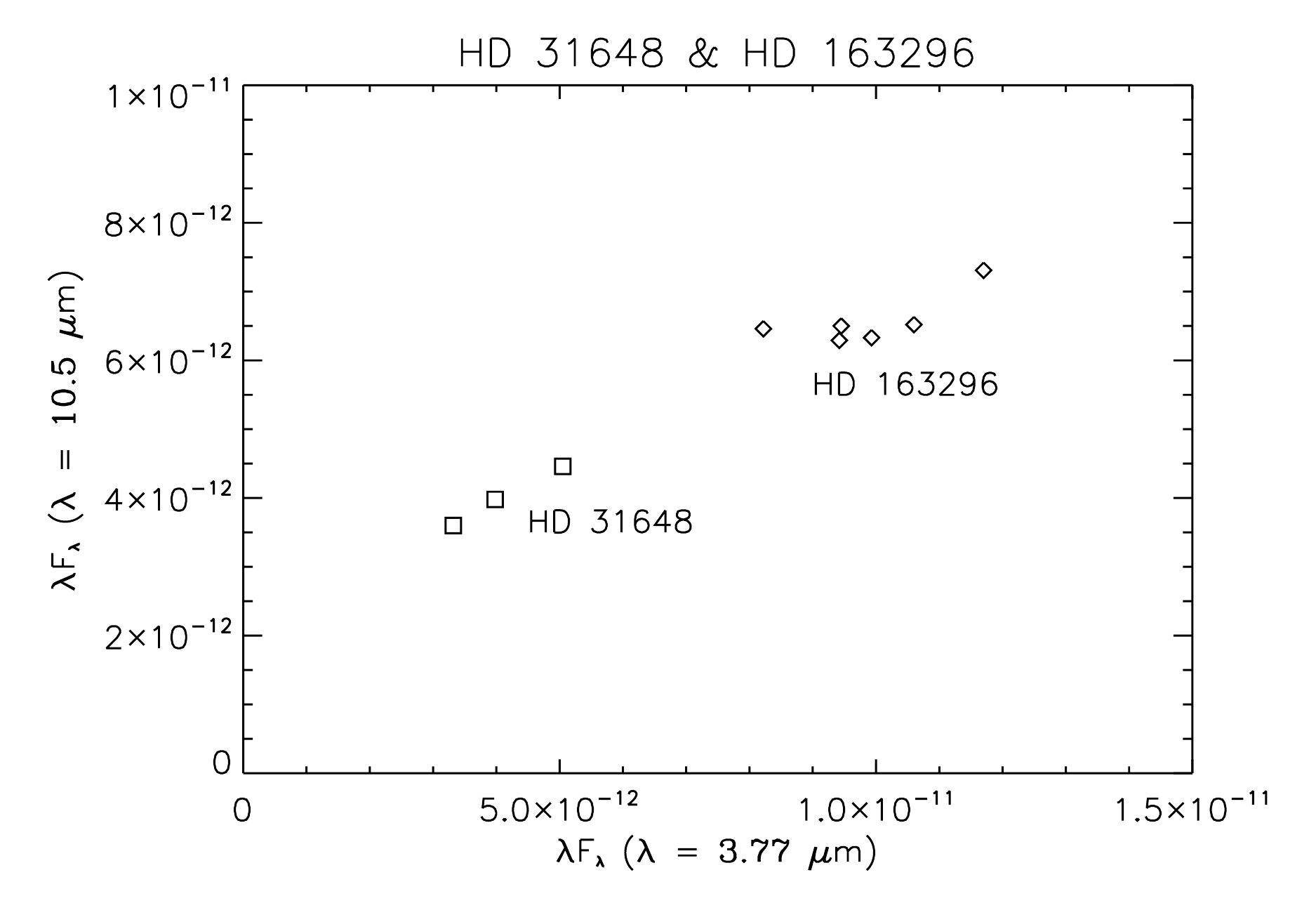}
\caption{Variability of HD 31648 and HD 163296 at 3.77 $\mu$m versus  versus 10.5 $\mu$m. With the exception of the 2002 outburst, the flux near 10 $\mu$m is relatively constant in HD 163296, even when the 3 $\mu$m emission changes. For HD 31648, changes at 3 $\mu$m have so far been accompanied by similar changes at 10 $\mu$m. \label{fig7}}
\end{figure}

\clearpage
\begin{figure}
 \center
 \includegraphics[scale=0.9]{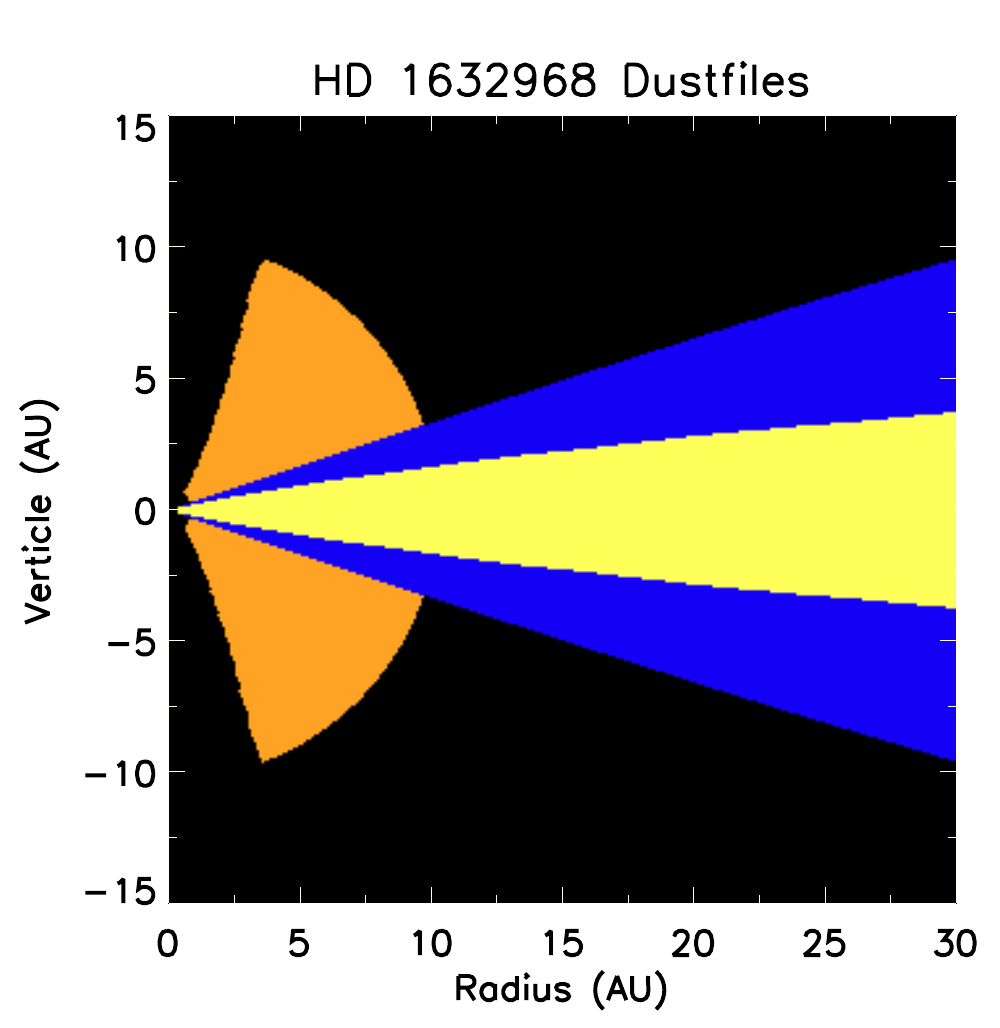}
\caption{The spatial distribution of the grain files used for the model of HD 163296. The envelope includes small ``ISM'' grains, while the disk contains the ``medium'' sized grains in its outer layers, and ``large'' grains closer to the disk mid-plane. The transition from the blue to yellow indicates where the density reaches the ``mid-plane'' regime (number density of H$_{2}$ $>$ 10$^{10}$ cm$^{-3}$). The model for HD 31648 was similar, except that the model fit was somewhat better without the transition to the largest grains in the mid-palne. \label{fig8}}
\end{figure}

\clearpage
\begin{figure}
 \center
 \includegraphics[scale=0.55]{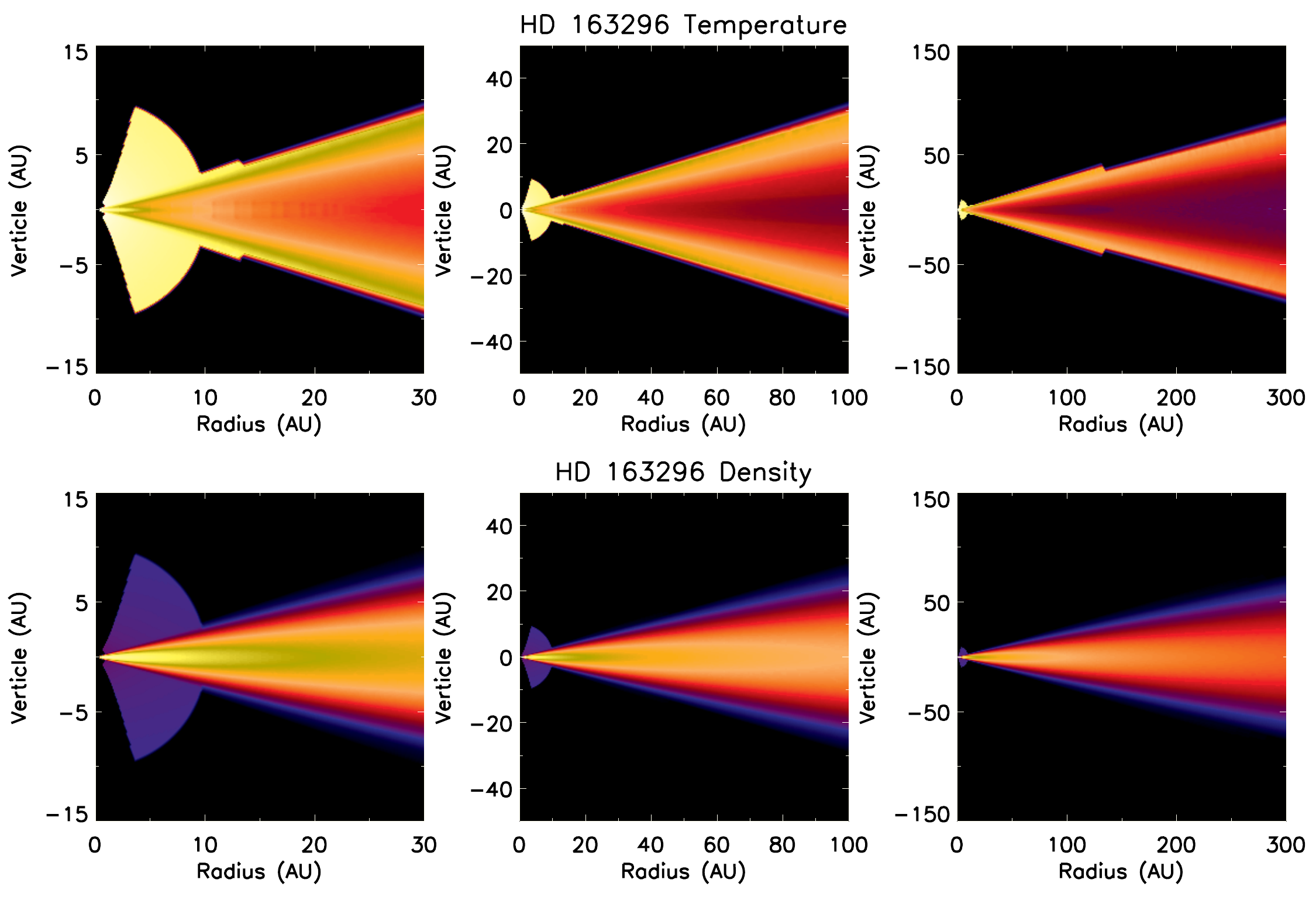}
  \center
   \includegraphics[scale=0.55]{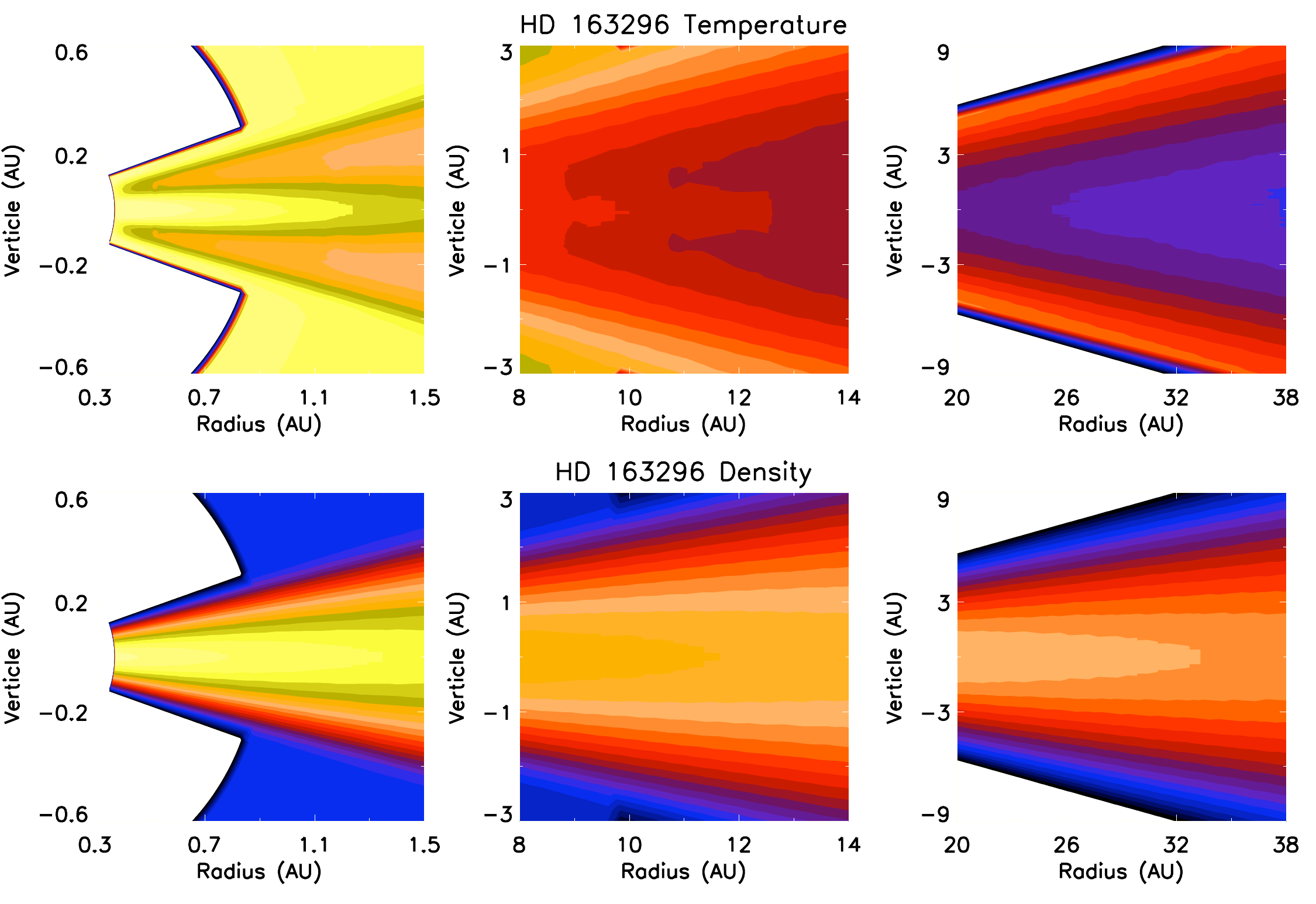}
\caption{The temperature and density structure for the model of HD 163296, at three different radial scales The color indicates how the temperature and density progress from lowest to highest values, with blue the lowest, followed by red, orange, yellow, and finally white for the highest. The ``truncated envelope'' has a density approximately ten orders of magnitude lower than the disk midplane, but receives direct illumination from the star.   \label{fig9}}
\end{figure}

\clearpage

\begin{figure}
 \center
 \includegraphics[scale=1.0]{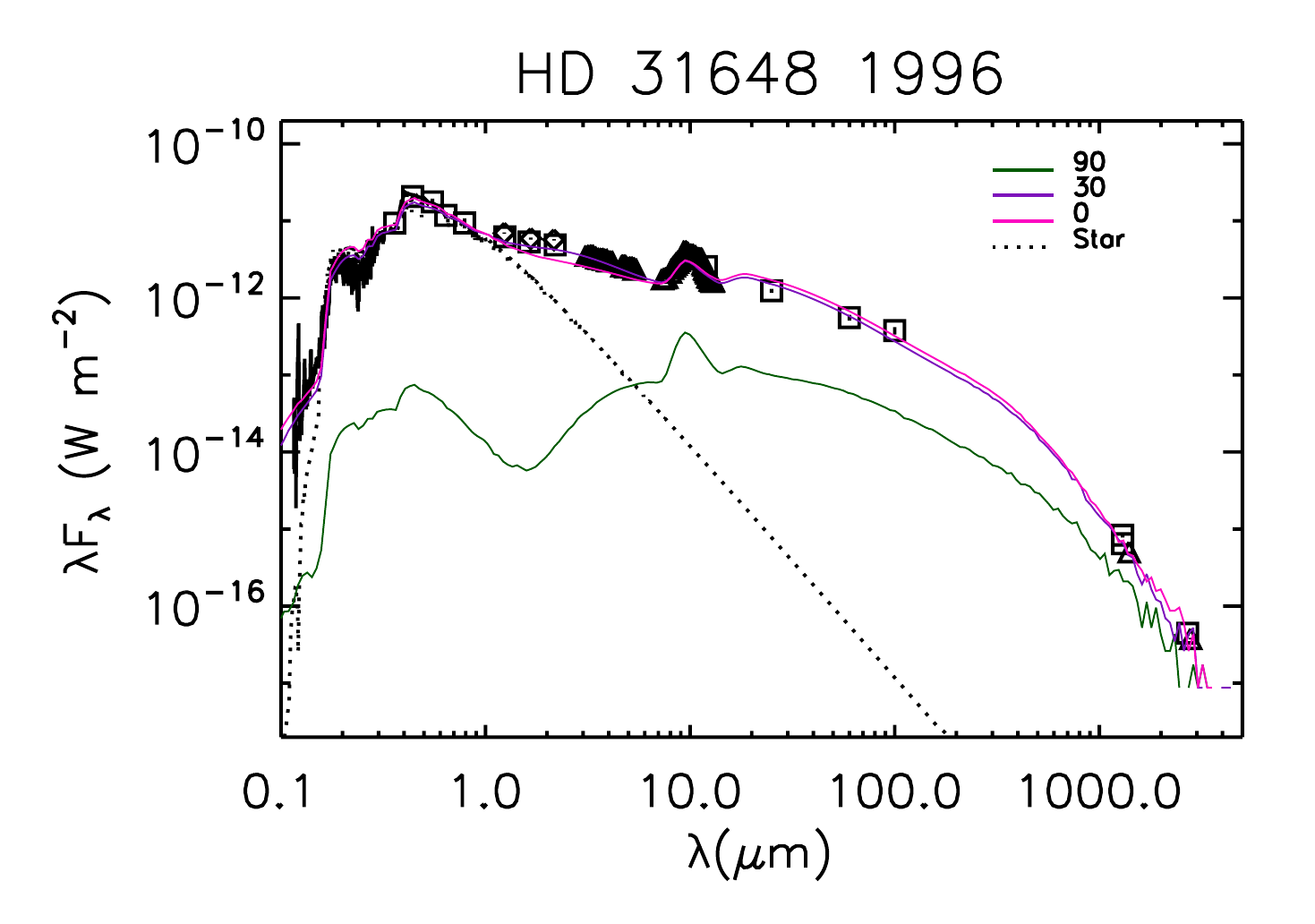}
\caption{A model of the IR emission of HD 31648 using the Monte Carlo radiative transfer code of \citet{whi03a,whi03b,whi04}, and described in the text . For this model the BASS spectrum from 1996 has been used. For simplicity, we show the model results for only three disk inclinations, 0$^{\circ}$ (upper solid line), 90$^{\circ}$ (lower solid line), and  30$^{\circ}$ (intermediate solid line). The last of these is close to the inclination values of \citet{sim01} and \citet{pi06}. For the stellar photospheric emission, a Kurucz model atmosphere with T$_{eff}$ = 8250 K and log g = 4.5 was used. \label{fig10}}
\end{figure}
 
 \clearpage

\begin{figure}
\center
 \includegraphics[scale=1.0]{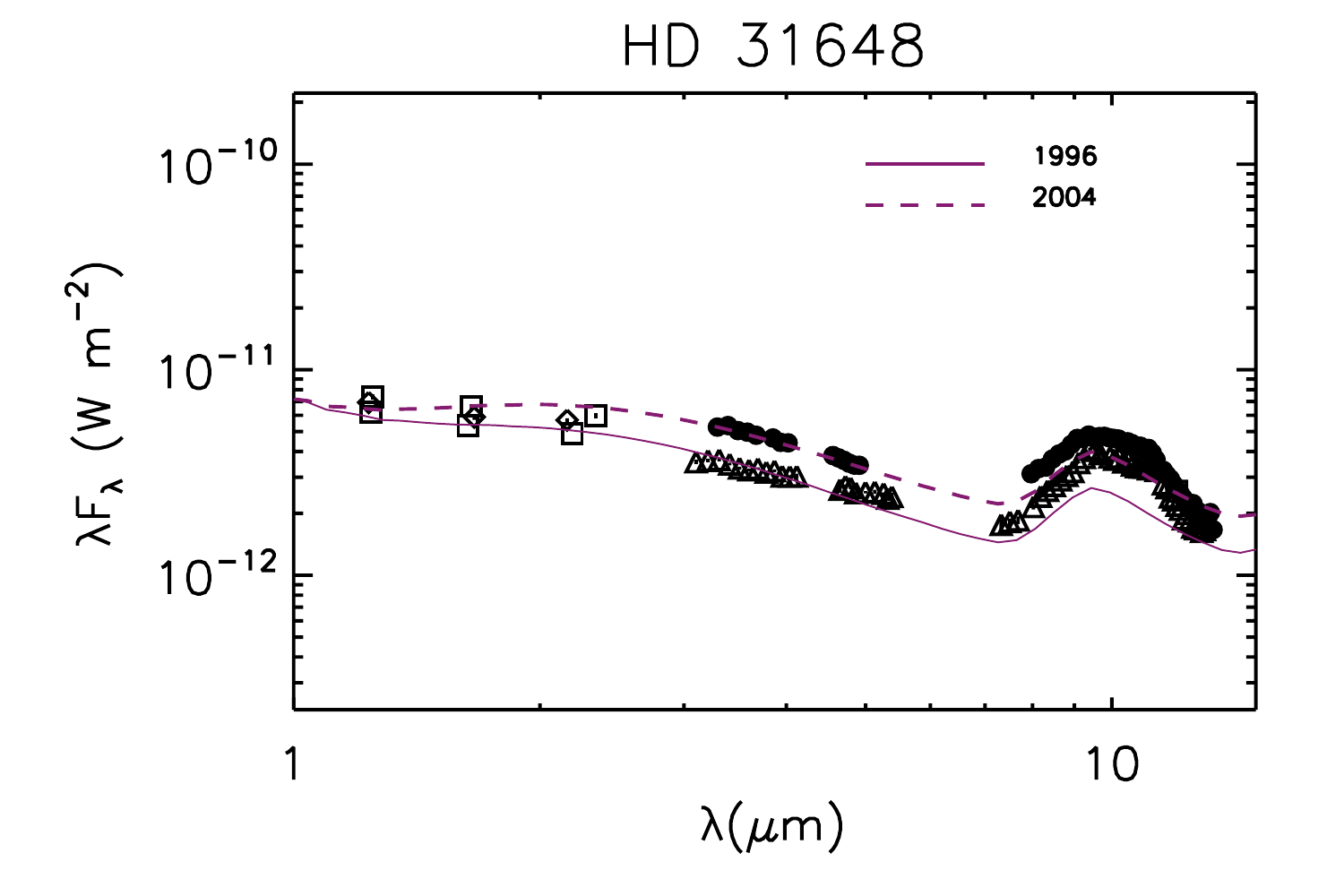}
\caption{Models for HD 31648 on two epochs. The 3-13 $\mu$m data in the triangles consists of the BASS spectrum from 1996, with shorter-wavelength 1998 observations of similar flux levels (lower open squares), from the EXPORT project \citep{eir01}.  The dotted line is the model for the 1996-98 data that is shown in Figure 10.  The shorter wavelength observations represented by diamonds were taken from the 2MASS catalog, and are also not contemporaneous with the BASS observations, but of approximately the same flux state. The filled circles are the BASS data from 2004. For the 2004 epoch, the fit was achieved with the same model as for the earlier epoch, except  the disk scale height was increased from 0.0076 AU to 0.0130 AU, and had a very slight increase in envelope mass. The upper squares are the JHK observations from Oct. 1980 at MLOF  (see Appendix A and Table 1), are not contemporaneous, but used to illustrate the general shape of the SED at a higher flux state. The dashed line is the same photospheric model used in Figure 10. These particular models slightly underpredict the strength of the silicate band, and suggests that the very surface layers of the disk may have grain sizes slightly smaller than those used here.  \label{fig11}}
\end{figure}
 
\clearpage

\begin{figure}
 \center
 \includegraphics[scale=1.0]{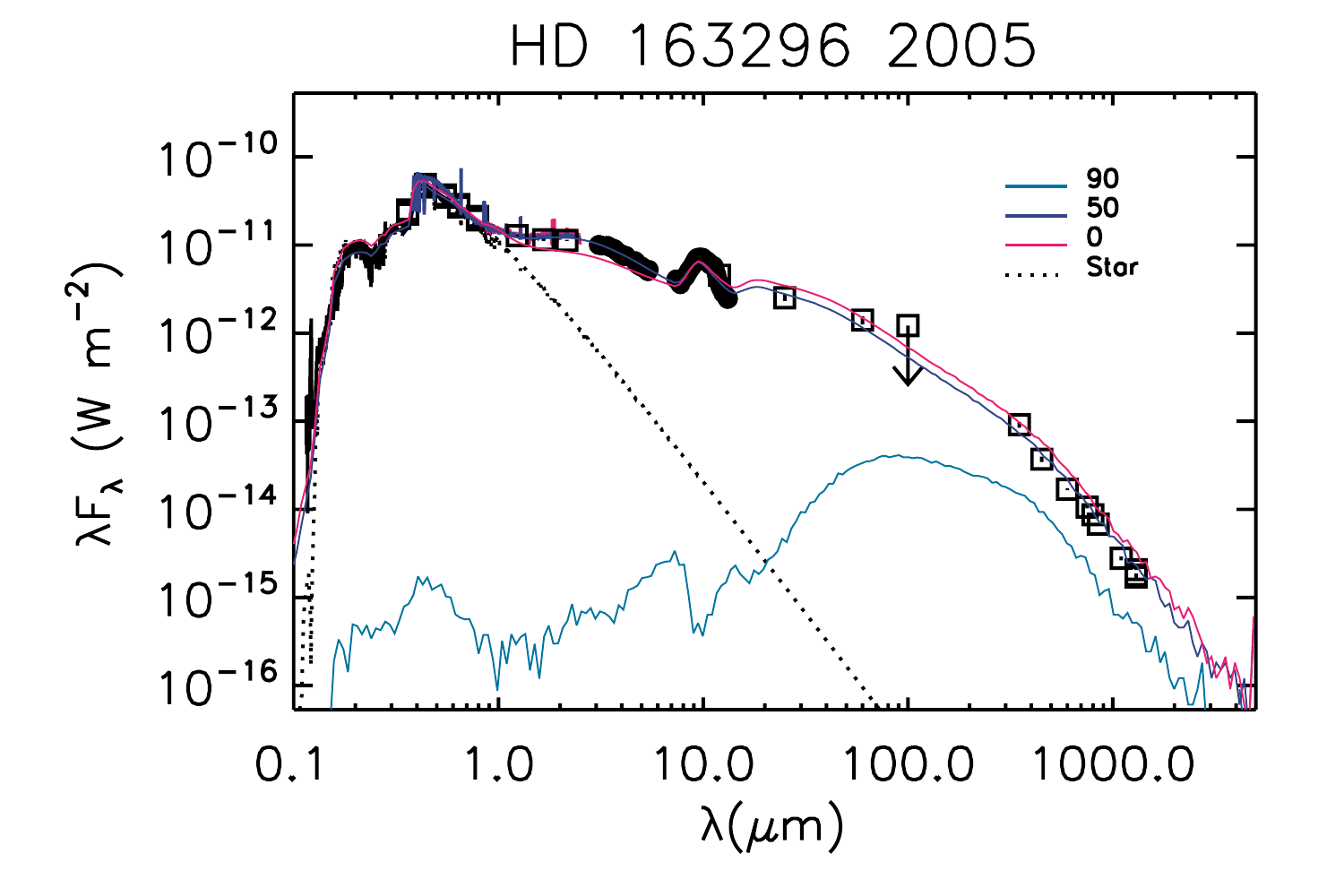}
\caption{A model of the IR emission of HD 163296 in 2005. The upper and lower solid lines are for inclinations of 0$^{\circ}$ and 90$^{\circ}$, respectively. The middle curve is  50$^{\circ}$, consistent, within the uncertainties, of the values reported in the literature \citep{ms97,gra00,was06,ise07}.  For the stellar photospheric emission, a Kurucz model atmosphere with T$_{eff}$ = 8750 K and  log g = 3.5 was used.\label{fig12}}
\end{figure}
 
\clearpage

\begin{figure}
 \center
 \includegraphics[scale=1.0]{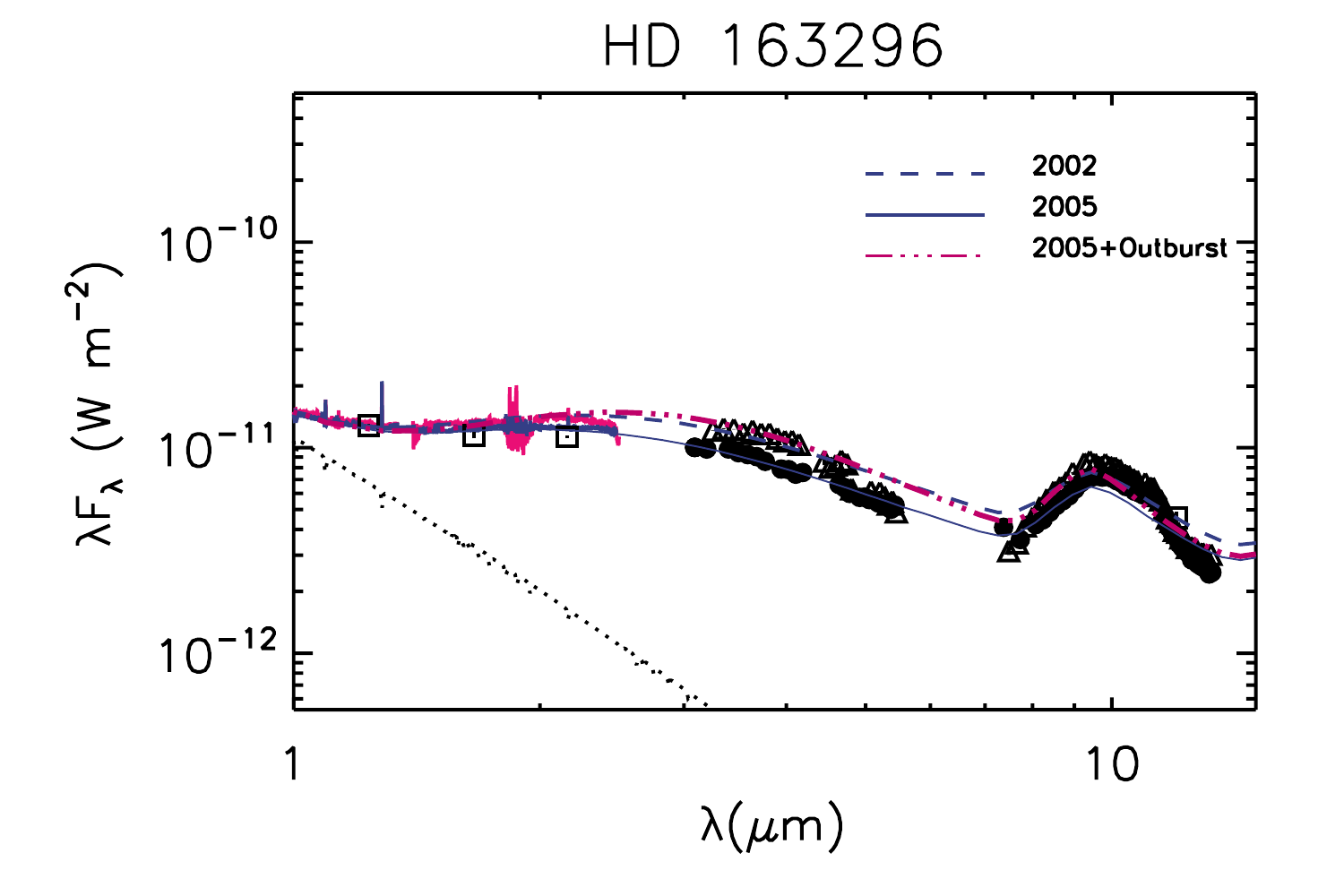}
\caption{A model of the IR emission of HD 163296 in 2002 using the same model as 2005, but increasing the disk scale height from 0.012 to 0.018 AU.  Also shown is the same model, with the added ``outburst'' component discussed in the text, and shown in Figures 4 and 14. The BASS and VNIRIS data are the same as those shown in Figure 4. \label{fig13}}
\end{figure}

\clearpage

\begin{figure}
 \center
 \includegraphics[scale=1.0]{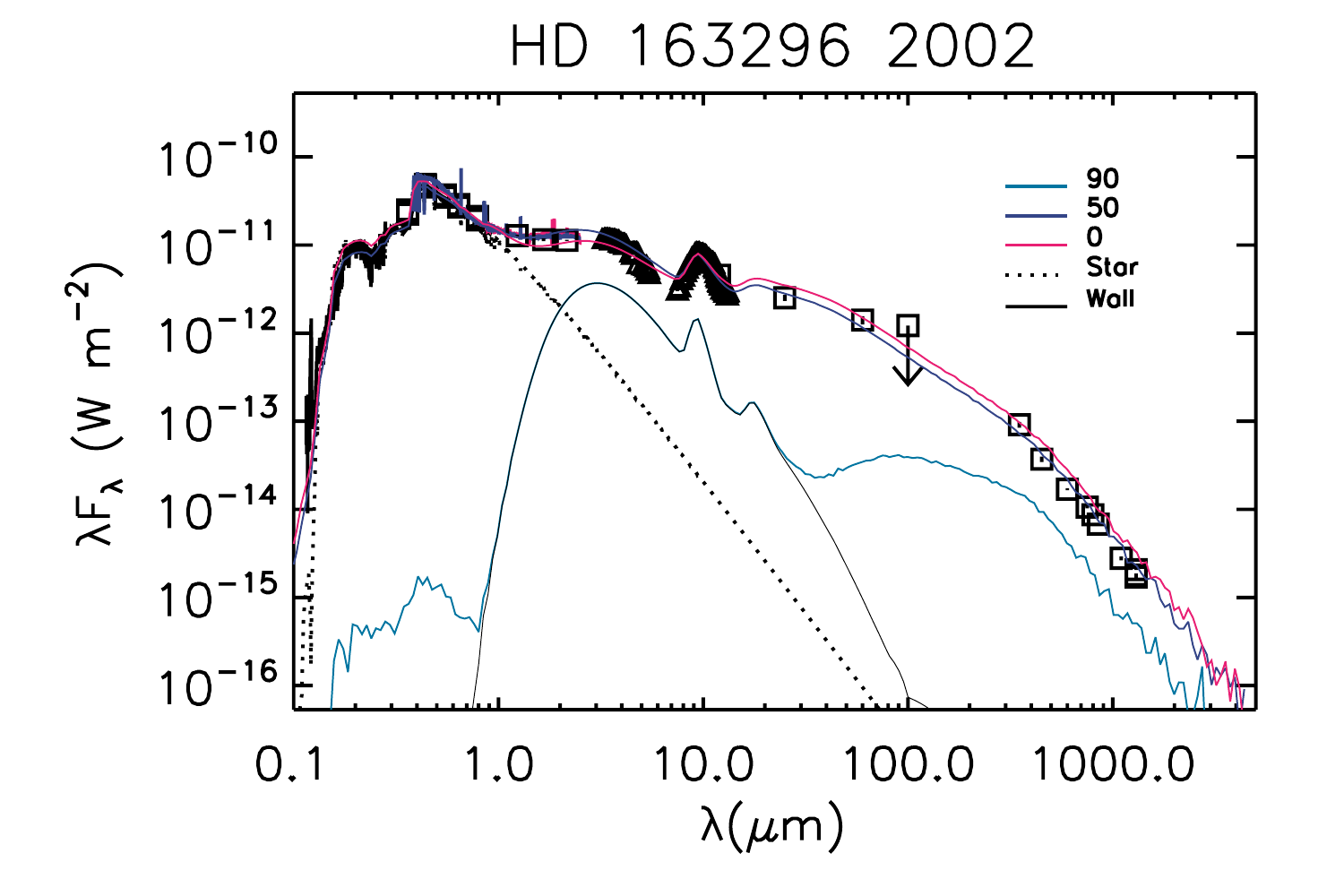}
\caption{A model of the IR emission of HD 163296 in 2002, using the additional component responsible for the outburst has been added to the 2005 fit. \label{fig14}}
\end{figure}
 
\clearpage

\begin{figure}
 \includegraphics[scale=0.45]{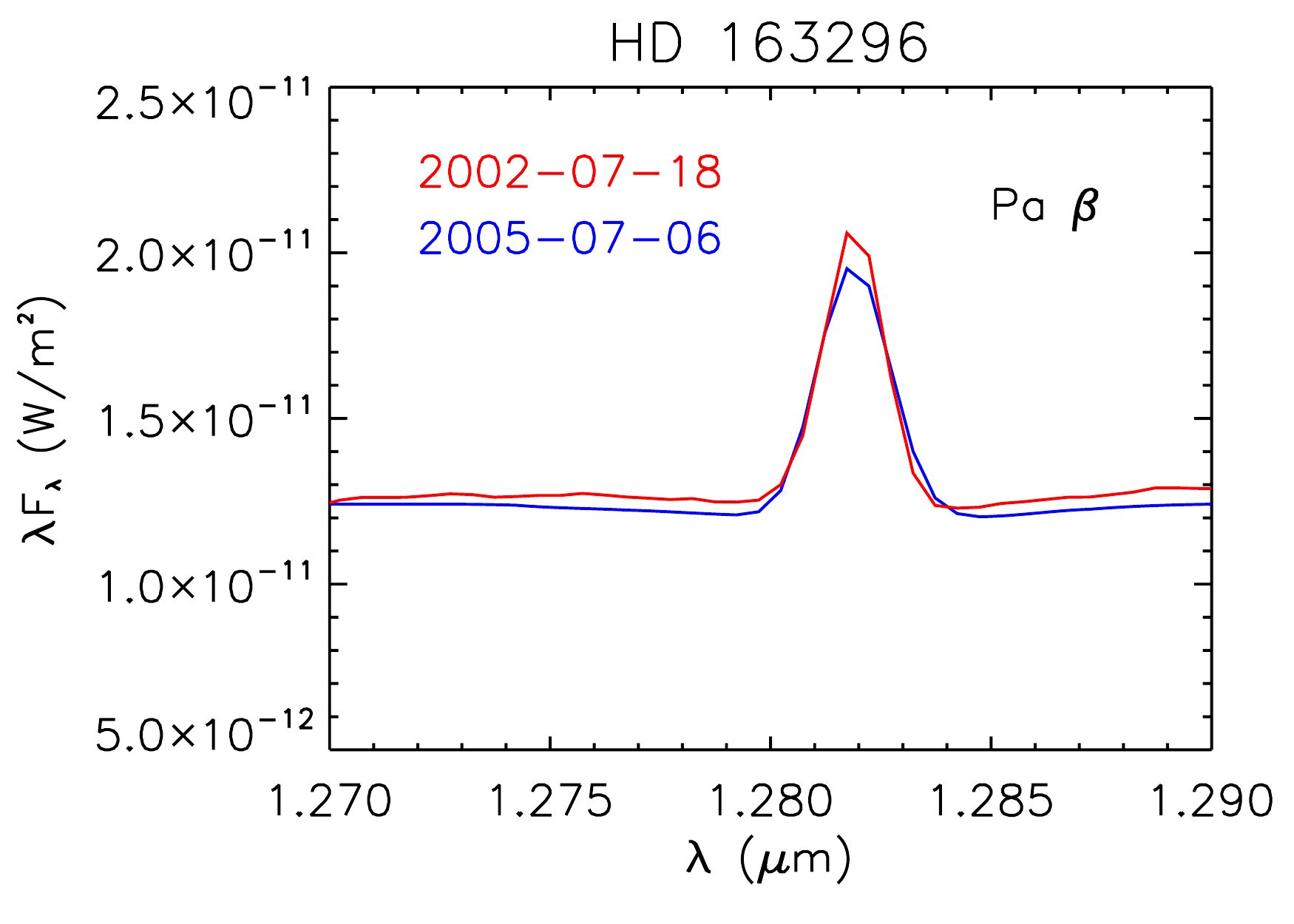}
  \includegraphics[scale=0.45]{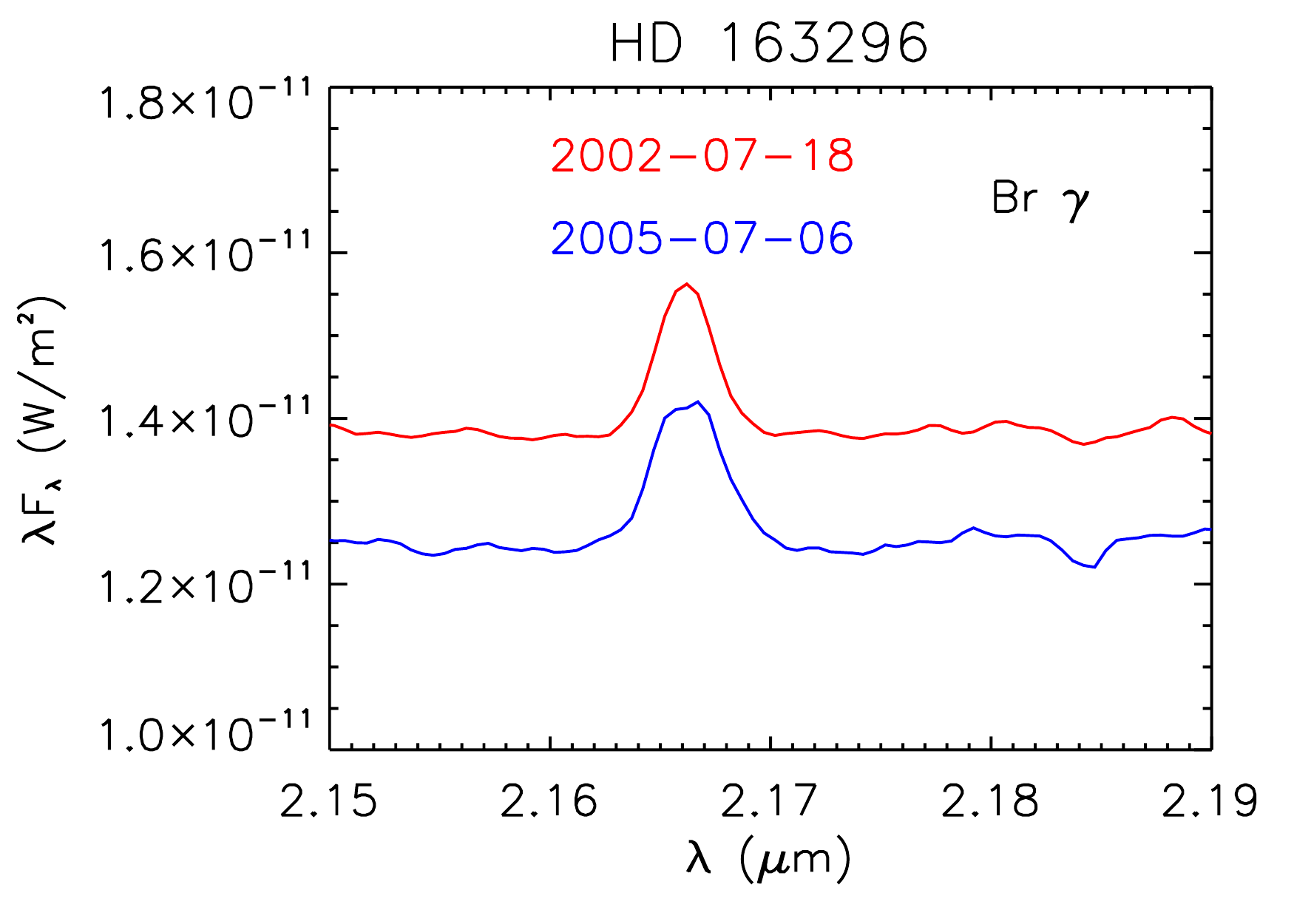}
   \center
   \includegraphics[scale=0.45]{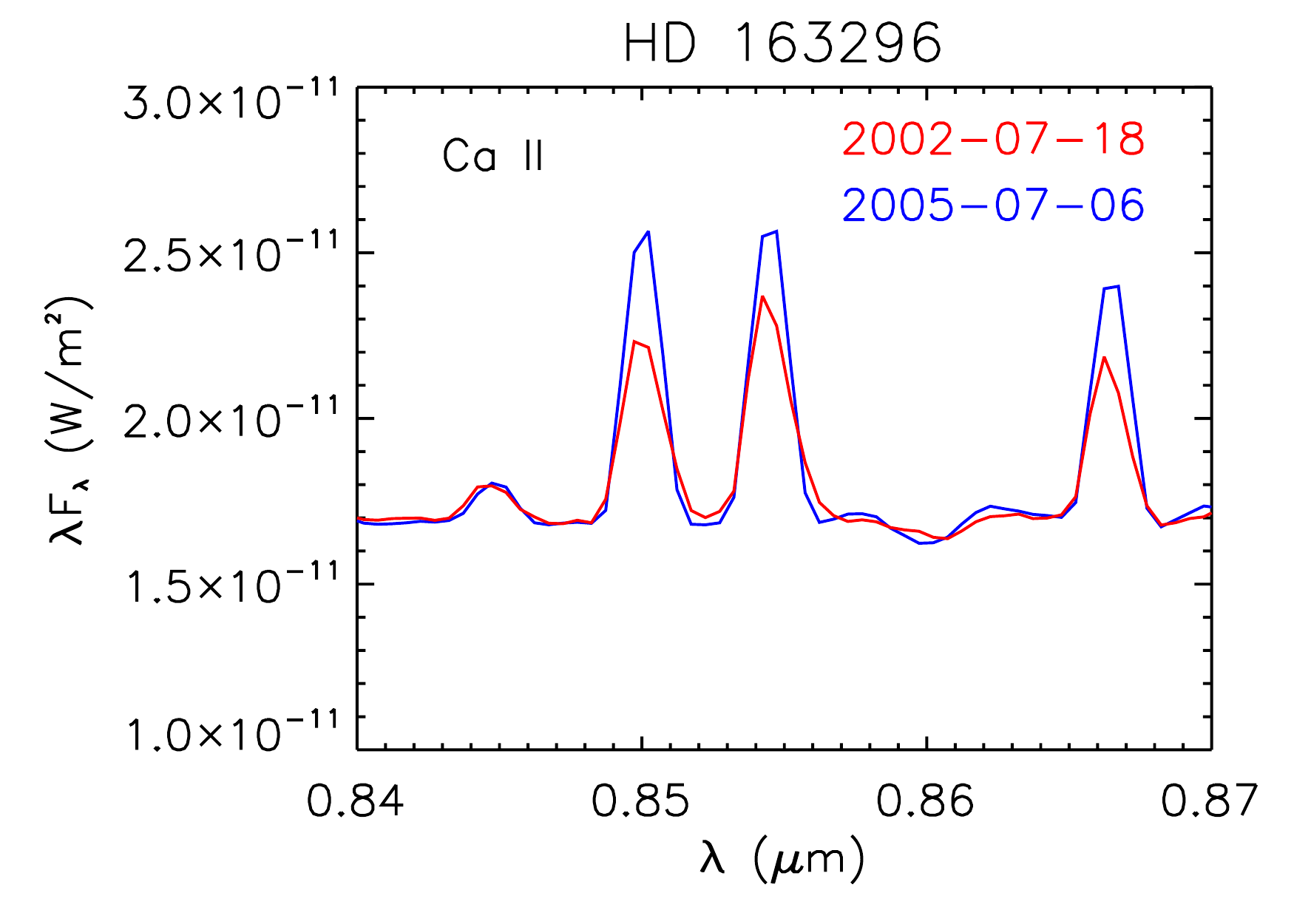}

\caption{The Ca II triplet, Pa $\beta$, and Br $\gamma$ lines on the two epochs for which VNIRIS data are available. The equivalent width of the Ca II feature at 0.850 $\mu$m was approximately 4.7 \AA \     in 2002 and 6.6 \AA  \   in 2005. The decreased strength in 2002, when the hump was strongest, suggests that the change may have been due to either increased occultation of the triplet-emitting region as the hump outburst proceeded, or by the addition of an unresolved absorption component in the line profiles. No significant changes were observed in the strengths of the hydrogen lines. \label{fig15}}
\end{figure}
 
 \clearpage

\begin{figure}
 \center
 \includegraphics[scale=0.9]{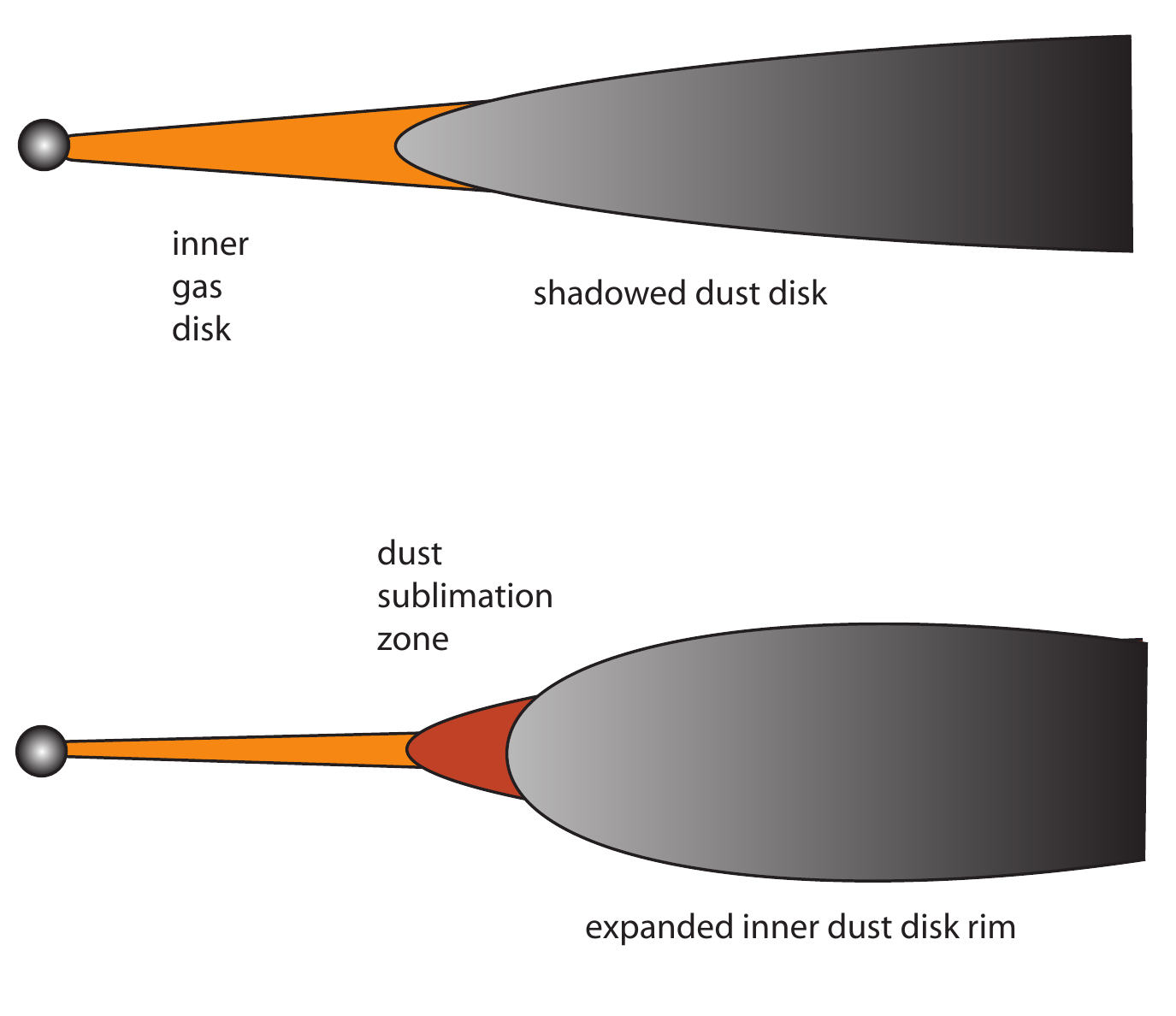}
\caption{Pictogram of possible changes occurring in the inner disk region of HD 163296, as described in the text.  The shapes of these regions are meant to be illustrative of one way that the structure might change, not an exact representation of the structure, since the current radiative transfer code does not include shadowing by the inner gas disk. \label{fig16}}
\end{figure}

\clearpage

\begin{figure}
 \center
 \includegraphics[scale=0.9]{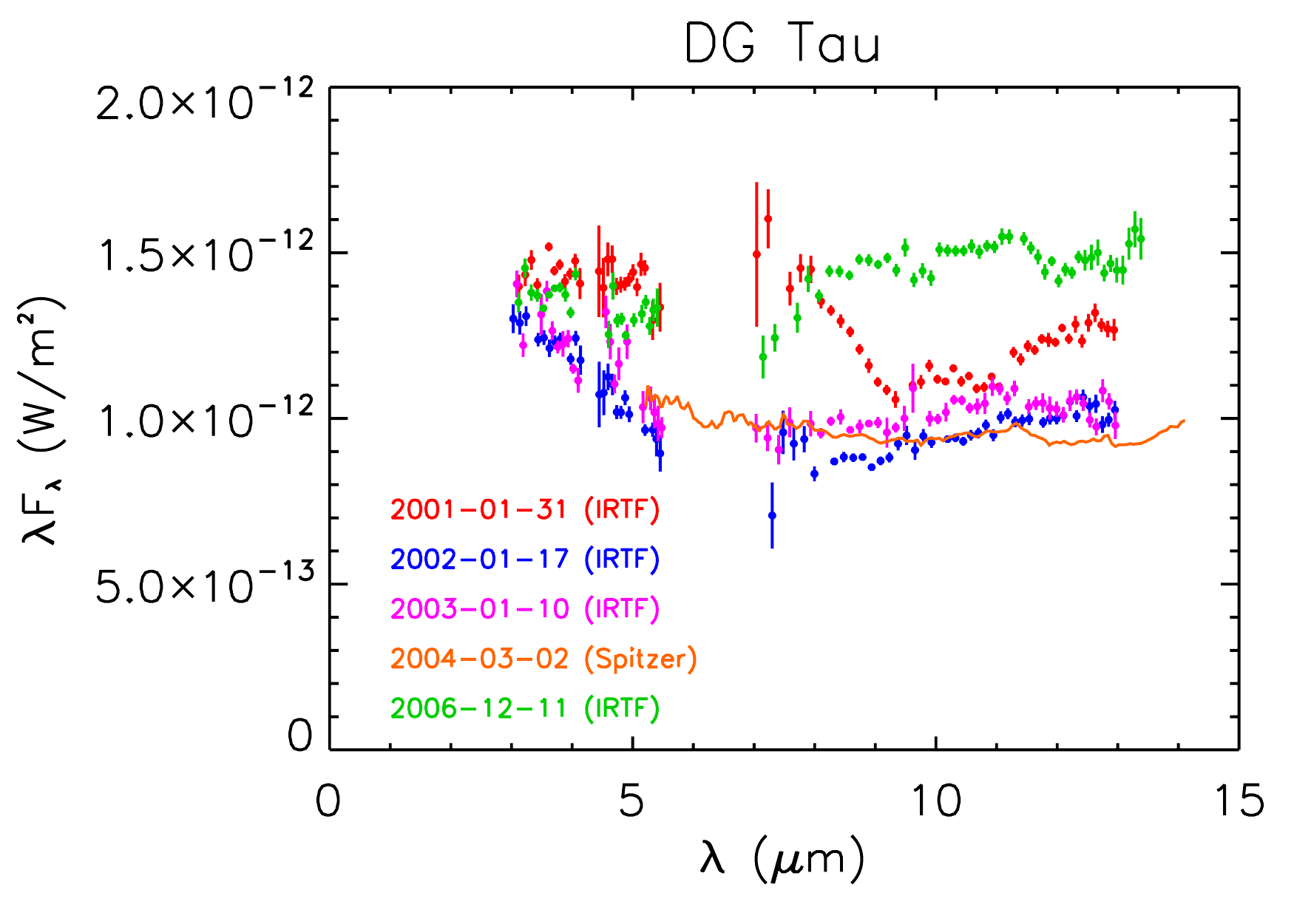}
\caption{DG Tau, observed on 4 dates between 2001 and 2006. Also shown is one spectrum obtained using the SL1 grating of the Infrared Spectrograph (IRS) of the Spitzer Space Telescope (from the SST Archive).  \label{fig17}}
\end{figure}
 
\clearpage

\begin{figure}
 \center
 \includegraphics[scale=0.9]{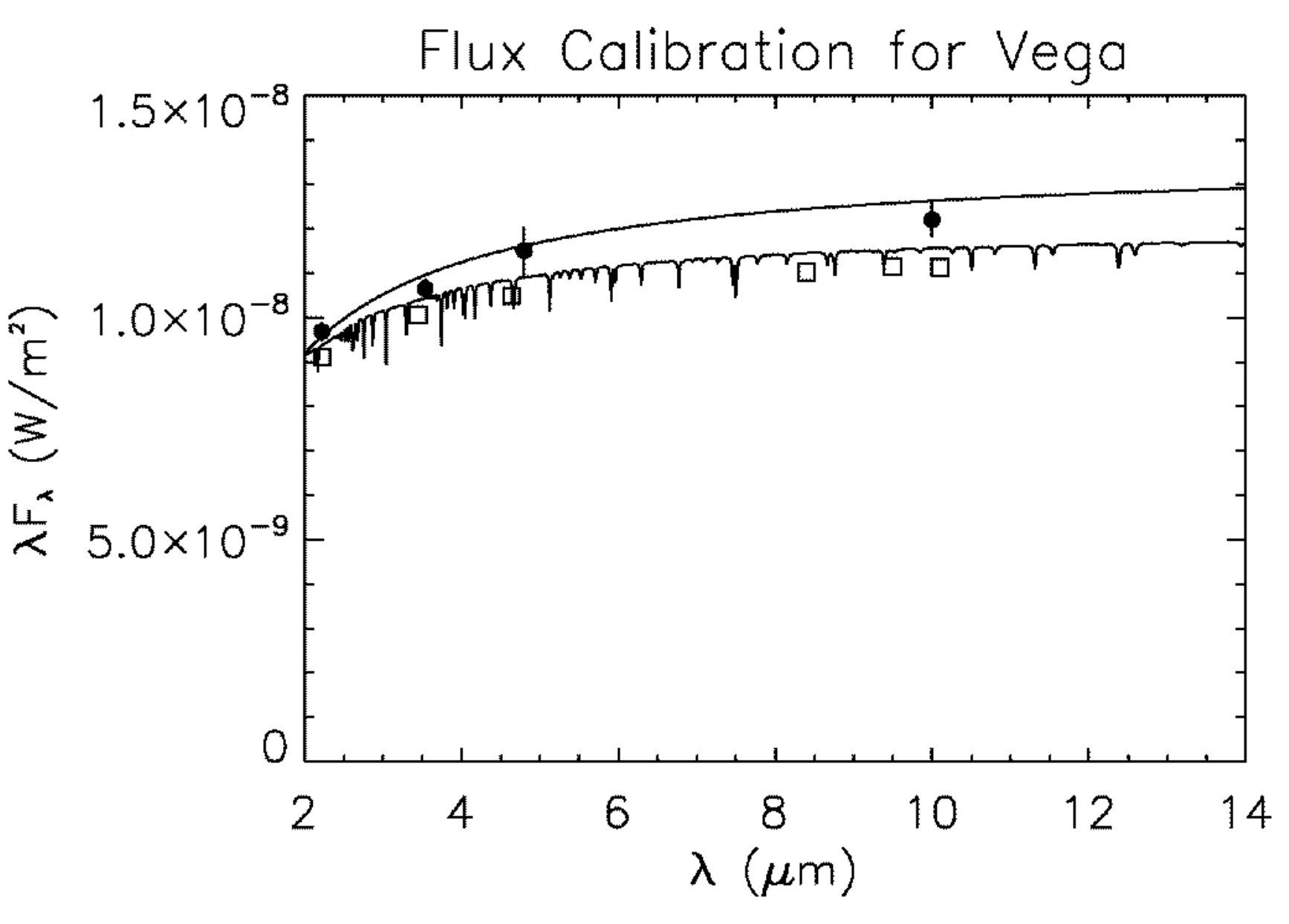}
\caption{A comparison of the various flux calibration systems. The data show the absolute flux of $\alpha$ Lyr for: BASS (upper curve), \citet{cohen92} (lower curve), earlier  \citet{sit81} (open squares), and empirical values from \citet{cam85} and \citet{rie85} (filled circles). \label{fig18}}
\end{figure}

\clearpage

\begin{deluxetable}{lrcc}
\tabletypesize{\scriptsize}
\tablecaption{Observations \label{tbl-1}}
\tablewidth{0pt}
\tablehead{
\colhead{Star} & 
\colhead{UT Date}  &
\colhead{Telescope} &
\colhead{Instrument} }  

\startdata
HD 31648 & 31 Aug 1979 & KPNO 1.3m & Bolo \\
  & 03 Jan 1980 & KPNO 1.3m & Bolo \\
  & 06 Oct 1980 & MLOF 1.5m & Uplooker3 \\
  & 08 Oct 1980 & MLOF 1.5m & Uplooker3 \\ 
  & 02 Dec 1984 & KPNO 1.3m & Hermann \\
  & 14 Oct 1996 & IRTF & BASS \\
  & 05 Aug 2004 & IRTF & BASS \\
  & 11 Dec 2006 & IRTF & BASS \\
  \\
HD 163296 & 1 Sep 1979 & KPNO 1.3m & Bolo \\
  & 02 Sep 1979 & KPNO 1.3m & Otto \\
  & 14 Oct 1996 & IRTF & BASS \\
  & 12 May 1998 & IRTF & BASS \\
  & 18 July 2002 & Lick 3m & NIRIS \\
  & 29 July 2002 & IRTF & BASS \\
  & 29 May 2004 & MLOF 1.5m & BASS \\
  & 06 July 2005 & Lick 3m & VNIRIS \\
  & 26 July 2005 & IRTF & BASS \\
  & 19 May 2006 & IRTF & BASS \\
 \enddata
\tablecomments{``Bolo'' refers to the bolometer of the Kitt Peak National Observatory (KPNO). The system utilized one ``flavor'' of JHKLMNQ broad-band filters, as well as intermediate-bandpass filters located in the 10 $\mu$m region, useful for determining the strength of the silicate emission band. ``Otto'' and ``Hermann'' at KPNO were InSb photometers with similar JHKLM filters. ``Uplooker3'' was yet another InSb system used at the Mount Lemmon Observing Facility (MLOF), which at that time was operated by the University of California at San Diego and the University of Minnesota, and is currently supported by The Aerospace Corporation and the University of Minnesota. For all of these observations, detailed records of the sky conditions have been preserved, and in many cases (i.e. the KPNO bolometer data) raw data output suitable for re-processing have survived the quarter-century of time since the data were obtained.}
\end{deluxetable}

\clearpage

\begin{deluxetable}{lcc}
\tabletypesize{\scriptsize}
\tablecaption{Model Parameters \label{tbl-1}}
\tablewidth{0pt}
\tablehead{
\colhead{Parameter} & 
\colhead{HD 31648}  &
\colhead{HD 163296} }  

\startdata
Star T$_{eff}$ (K) &  8250 & 8750 \\
Distance (pc) & 122 & 140 \\
A$_{V}$ (mag) & 0.05 & 0.10 \\	 
Disk Midplane Dust File\tablenotemark{a} &	Medium & Large \\
Disk Dust File\tablenotemark{a} & Medium & Small \\
Envelope Dust File\tablenotemark{a} & Small & Small \\
Disk Outer Radius (AU) & 250  & 450 \\
Disk Inner Radius (AU) & 0.196/0.245 & 0.29/0.35 \\
Disk Scale Height  - Inner (AU)\tablenotemark{b} & 0.0076/0.0130 & 0.012/0.018 \\
A - Disk Mass Density Exponent\tablenotemark{c} & 1.6 & 1.99 \\
B - Disk Scale Height Exponent\tablenotemark{d} & 0.6 & 0.99 \\
Disk Accretion Rate (M$_{\sun}$  yr$^{-1}$) & 4.0x10$^{-9}$ & 8.0x10$^{-8}$\\
Disk Mass (M$_{\sun}$) & 0.8 & 0.4 \\
Envelope Max Radius (AU) & 10 & 10 \\
Envelope Min Radius (AU) & 0.24 & 0.83 \\
Inner Cavity Wall Opening Angle (deg) & 10 & 20 \\
Envelope Mass Infall Rate (M$_{\sun}$  yr$^{-1}$)\tablenotemark{c} & 7.0x10$^{-7}$/1.2x10$^{-6}$ & 7.0x10$^{-7}$ \\
Envelope Mass (M$_{\sun}$ yr$^{-1}$)\tablenotemark{c} & 4.0x10$^{-8}$/6.4x10$^{-8}$ & 2.6x10$^{-8}$ \\
Accretion Shock Luminosity Fraction & 1.1\% & 11.0\% \\
C - Surface Brightness Exponent\tablenotemark{e} & -3.1 & -3.1 \\
\enddata
\tablecomments{The sublimation temperature for all the models was set to be 1600 K. According to \citet{gail02}, for disk mid-pane conditions, the sublimation/condensation temperatures for enstatite (MgSiO$_{3}$) and forsterite (Mg$_{2}$SiO$_{4}$) silicates are between 1300 and 1500 K. For some other minerals that could provide hot dust emission, like perovskite (CaTiO$_{3}$), this temperature can be significantly higher. }
\tablenotetext{a}{The ``small'', ``medium'', and ``large'' grains refer to the interstellar medium grain model of \citet{kmh94}, the HH30 scattered light grain model of  \citet{cot01}, and the HH30 disk thermal emission model of  \citet{wood02}, respectively.}
\tablenotetext{b}{A range in values are for the ``low'' and ``high'' flux states.}
\tablenotetext{c}{Describes the radial mass density distribution within the disk.: $\rho \propto R^{-A}$}
\tablenotetext{d}{Describes the change in scale height of the disk with radial distance: $H \propto R^{B}$. B = 1 corresponds to a disk with a conical opening. B $>$ 1 indicates a concave upward flared disk, while B $<$ 1 is for a convex upward ``anti-flared'' disk.}
\tablenotetext{e}{The predicted $\lambda$=1.14$\mu$m surface brightness distribution, where the surface brightness  $\propto$R$^{-C}$ . The value listed for HD 163296 matches the observed value, based on coronagraphic imaging with the \textit{Hubble Space Telescope}. Although not detected in the earlier survey of \citet{gra05}, the disk of HD 31648 has recently been detected \citep{ste07}, and is steeper than predicted by the current model.}

\end{deluxetable}

\end{document}